\documentclass[prd,superscriptaddress,showpacs,amssymb,amsmath,amsfonts,aps,altaffilletter,nofootinbib,letterpaper,twocolumn]{revtex4-1}

\usepackage{amssymb}
\usepackage{color}
\usepackage{graphicx}
\usepackage{fancyhdr}
\usepackage{float}
\usepackage[raggedright]{subfigure}
\usepackage{multirow}
\usepackage{array}
\usepackage{hyperref}
\usepackage{bm}
\usepackage[normalem]{ulem}

\definecolor {darkgreen}{rgb}{0.2, 0.7, 0.2}

\renewcommand{\vec}[1]{\mathbf{#1}}

\newcommand{\Msun}{\mathrm{M}_\odot}
\newcommand{\mtrx}[1]{\mathbf{#1}}
\renewcommand{\th}{\mathrm{th}}
\newcommand{\avg}[1]{\left<#1\right>}
\newcommand{\ip}[2]{\left< #1,\, #2 \right>}
\newcommand{\arbip}[2]{\left<\left< #1,\, #2 \right>\right>}
\renewcommand{\d}{\mathrm{d}}
\newcommand{\extP}{\vec{\Xi}}
\newcommand{\intP}{\vec{\Upsilon}}
\newcommand{\allP}{\vec{\Theta}}
\newcommand{\pls}{+}
\newcommand{\crs}{\times}
\newcommand{\rhonew}{\tilde{\rho}}%_{\mathrm{new}}}
\newcommand{\eff}[2]{\mathcal{E}_{#1 #2}}
\newcommand{\effcy}[2]{\epsilon_{#1 #2}}
\newcommand{\dm}{\mathrm{D}}
\newcommand{\sd}{\mathrm{S}}
\newcommand{\effdist}{\mathcal{D}}
\newcommand{\vol}[2]{V_{#1 #2}}
\newcommand{\gain}[2]{\mathcal{G}^{#1}_{#2}}
\newcommand{\netgain}[2]{\mathrm{net}\left\{\mathcal{G}^{#1}_{#2}\right\}}

\newcommand{\pdf}[1]{f_{#1}}
\newcommand{\cdf}[1]{F_{#1}}

\newcommand{\RatesPaper}{Abadie:2010cf}
\newcommand{\sSixLowMass}{Abadie:2011np}
\newcommand{\sSixHighMass}{Aasi:2012}
\newcommand{\sFiveHighMass}{Abadie:2011}
\newcommand{\LigoSearchPapers}{\sSixLowMass, \sSixHighMass}

\begin{document}

\title{Impact of Higher Harmonics in Searching for Gravitational Waves from Non-Spinning 
Binary Black Holes}

\author{Collin Capano} \author{Yi Pan} \author{Alessandra Buonanno}
\affiliation{Maryland Center for Fundamental Physics \& Joint Space
  Science Institute, Department of Physics, University of Maryland,
  College Park, 20742}

\begin{abstract}

Current searches for gravitational waves from coalescing binary black holes
(BBH) use templates that only include the dominant harmonic.  In this study we
use effective-one-body multipolar waveforms calibrated to numerical-relativity
simulations to quantify the effect of neglecting sub-dominant harmonics on the
sensitivity of searches. We consider both signal-to-noise ratio (SNR) and the
signal-based vetoes that are used to re-weight SNR. We find that neglecting
sub-dominant modes when searching for non-spinning BBHs with component masses
$3\,\Msun \leq m_1, m_2 \leq 200\,\Msun$ and total mass $M < 360\,\Msun$ in
advanced LIGO results in a negligible reduction of the re-weighted SNR at
detection thresholds.  Sub-dominant modes therefore have no effect on the
detection rates predicted for advanced LIGO.  Furthermore, we find that if
sub-dominant modes are included in templates the sensitivity of the search
becomes \emph{worse} if we use current search priors, due to an increase in
false alarm probability.  Templates would need to be weighted differently than
what is currently done to compensate for the increase in false alarms.  If we
split the template bank such that sub-dominant modes are only used when $M
\gtrsim 100\,\Msun$ and mass ratio $q \gtrsim 4$, we find that the sensitivity
does improve for these intermediate mass-ratio BBHs, but the sensitive volume
associated with these systems is still small compared to equal-mass systems.
Using sub-dominant modes is therefore unlikely to substantially increase the
probability of detecting gravitational waves from non-spinning BBH signals
unless there is a relatively large population of intermediate mass-ratio BBHs
in the universe.

\end{abstract}

\maketitle

\section{Introduction}

Within the next few years the next generation of gravitational-wave detectors
will come online. These detectors --- the advanced Laser Interferometer
Gravitational-wave Observatory (aLIGO) in the United States
\cite{Harry:2010zz}, the French-Italian Virgo observatory \cite{AdVTDR}, the
KAGRA detector in Japan \cite{Somiya:2011np}, and a potential third LIGO
detector in India \cite{INDIGOwebsite} --- will be sensitive to sources up to
$10$ times more distant than first generation detectors. One of the most
promising sources of gravitational-waves for these detectors are coalescing
binary black holes (BBHs). As the two black holes in a binary orbit each other,
they emit gravitational radiation; this causes them to inspiral and eventually
merge into a single black hole.  Advanced detectors will be able to detect
radiation emitted during this process up to tens of Gpc away.
The rate of BBH coalescences with masses detectable by the advanced detectors
is highly uncertain: the detection rate has been estimated to be as low as
$0.4\,\mathrm{yr}^{-1}$, but it may be as high as $1000\,\mathrm{yr}^{-1}$
\cite{\RatesPaper}. If the optimistic end of this range is correct, BBHs would
be the most prolific source of gravitational-wave detections in the advanced
detector era.

Searches for BBHs use a matched filter to determine when a gravitational wave
is likely to be present in a detector's data
\cite{\LigoSearchPapers}.\footnote{Un-modeled ``burst" searches
\cite{Abadie:2012ab} are also sensitive to BBHs \cite{Pankow:2009lv}.
Previously, these searches have been used to search for signals with total
masses $M > 100\,\Msun$, as the signals in this mass range did not have many
cycles in the sensitive band of the initial generation of LIGO detectors. In
this paper, we will consider signals with $M$ as large as $360\,\Msun$.  Due to
the improved low-frequency ($<40\,$Hz) sensitivity of the advanced detectors,
template-based searches similar to those done in \cite{\LigoSearchPapers} will
be feasible at these higher masses.  We therefore only consider template-based
searches here.} The filter produces a signal-to-noise ratio (SNR) that is
proportional to the probability that a signal exists in the data with similar
parameters as the template used in the correlation \cite{Sathyaprakash:1991mt}.
Since the physical parameters of the signal are unknown \textit{a priori}, a
discrete \emph{bank} of templates is used to cover the range of possible
parameters \cite{Balasubramanian:1995bm, Owen:1995tm}.  This technique for
searching for gravitational waves relies on good agreement between templates
and real signals. If there is significant disagreement between the two, the SNR
will be reduced, and it becomes difficult to separate potential signals from
noise. It is therefore important to verify that templates adequately resemble
potential gravitational-wave signals.

Gravitational waveforms are decomposed into a $-2$ spin-weighted spherical
harmonic basis ${}_{-2}Y_{lm}(\theta, \phi)$. Template waveforms used in past
searches for BBHs have only included the most dominant mode, $l = |m| = 2$
\cite{\LigoSearchPapers}.  In addition, predictions of the number of detections
that will be made in advanced LIGO have been made assuming both templates and
signals only have the dominant mode \cite{\RatesPaper}. Real signals, however,
will have all modes. There can be significant \emph{mismatch} [defined as
$1-\eff{}{}$, where $\eff{}{}$ is given by Eq. \ref{eqn:defEff}; see Sec.
\ref{sssec:effectualness} for details] between waveforms that include
sub-dominant modes and waveforms that only include the dominant mode
\cite{Pan:2011gk, Pekowsky:2012sr, Brown:2013hm, McKechan:2011ps}. This raises
the question, by neglecting sub-dominant modes has the sensitivity of BBH
searches been overstated?  Furthermore, would sensitivity improve if templates
included sub-dominant modes?

Several studies have investigated the effects of sub-dominant modes to try to
answer these questions. Ref. \cite{Pekowsky:2012sr} used waveforms generated
from numerical relativity as both template and signal to measure mismatch when
sub-dominant modes are not included in templates. They considered systems with
total mass $M > 100\,\Msun$ and mass ratios $1 \leq q \leq 4$.\footnote{In this
paper we use the convention that $q = m_1/m_2$ with $m_1 \geq m_2$.} When
computing overlaps they only considered templates which had the same intrinsic
parameters (mass and spin) as signals.  They found that adding sub-dominant
modes could improve detection volume by up to $30\%$, but they noted that
the largest gain in volume occurred for signals for which the detectors had the
least sensitivity (these were systems with asymmetric masses and inclination
angle --- the angle between the orbital angular momentum and the line of sight
to the detector --- $\theta \approx \pi/2$).

To fully understand the effect of sub-dominant modes on the sensitivity of BBH
searches, a bank of template waveforms covering the source parameter space is
necessary. These waveforms should contain the complete inspiral, merger and
ringodwn stages of sub-dominant modes since the latter become relatively
stronger during the last stage of inspiral and merger. Here we employ
effective-one-body (EOB) waveforms~\cite{Buonanno:1998gg, Buonanno:2000ef}
calibrated to numerical-relativity simulations~\cite{Pan:2011gk}.

Building on an initial attempt~\cite{Buonanno:2007} to calibrate sub-dominant
modes, Ref.~\cite{Pan:2011gk} built a nonspinning EOB model that include four
sub-dominant modes, namely the $(l,m)=(2,1)$, $(3,3)$, $(4,4)$ and $(5,5)$
modes, as well as the dominant $(l,m)=(2,2)$ mode. The EOB model of
Ref.~\cite{Pan:2011gk} was calibrated to numerical-relativity simulations of
mass ratios $q\le6$.\,\footnote{An EOB model with slightly different
parametrization was calibrated to the same set of numerical-relativity
waveforms and provides two sub-dominant modes $(l,m)=(2,1)$ and
$(3,3)$~\cite{Damour:2012ky}.} A direct study~\cite{Littenberg:2012uj} carried
out with the Markov Chain Monte Carlo technique demonstrated that the EOB
waveforms of Ref.~\cite{Pan:2011gk} are indistinguishable from the
numerical-relativity waveforms~\cite{Buchman:2012dw} used to calibrate them up
to SNR = 50 for the advanced LIGO detectors. Furthermore, preliminary studies
in Ref.~\cite{Pan:2011gk} suggested that the EOB waveforms containing
subdominat modes could be sufficiently accurate to search for nonspinning BBH
signals of $q\le6$, at least in the relatively narrow frequency band where
direct comparison with numerical-relativity waveforms was possible. The recent
investigation of Ref.~\cite{Pan:2013len} verified the accuracy of these EOB
waveforms in the entire sensitivity band of advanced LIGO detectors. Having the
correct limit for $q\rightarrow\infty$ by construction, we expect EOB waveforms
to be reasonably accurate when $q>6$. This expectation was recently reinforced
by the excellent agreement found against a $q=10$ numerical-relativity
waveform~\cite{Hinder:2013oqa}. Finally, the dominant mode EOB waveforms that
we use here have been employed as simulated signals in the most recent LIGO BBH
searches~\cite{Aasi:2012}.

Using a bank of dominant-mode templates that allows maximization over the
masses of the binaries, Ref. \cite{Brown:2013hm} studied templates and signals
with component masses $3\,\Msun \leq m_1, m_2 \leq 25\,\Msun$. They also found
that sub-dominant modes had little effect on equal-mass systems, but argued
that sensitive volume could be increased by as much as $25\%$ for systems with
$q \geq 4$ and inclination angles $1.08\,\mathrm{rad} \leq \theta \leq
2.02\,\mathrm{rad}$ if sub-dominant modes were added to templates.

Both of these studies calculated what percentage of sensitive volume could be
gained if sub-dominant modes were added to templates by finding the fractional
loss in SNR of signals with sub-dominant modes when they were recovered by
dominant-mode templates.  However, estimating the gain in sensitive volume in
this manner neglects additional complications that arise when searching in
real, non-Gaussian, data. In real data a signal-based veto, $\chi^2$
\cite{Allen:2004}, is used to re-weight SNR \cite{\LigoSearchPapers}.
\emph{Re-weighted SNR} is needed in order to separate potential
gravitational-wave signals from non-Gaussian transients that are present in
detector data \cite{ihopePaper:2012}. Any mismatch between the templates and
signals causes an increase in $\chi^2$, which in turn causes a decrease in
re-weighted SNR relative to SNR. Thus the sensitivity of dominant-mode
templates to real signals may be worse than predicted from SNR considerations
alone. This makes the case for adding sub-dominant modes to templates stronger.

A signal must have high statistical significance in order for it to be
considered a gravitational-wave candidate. The standard measurement of
significance is \emph{false-alarm probability}, which is the probability that
an event caused by noise is mis-identified as a signal. When calculating
sensitive volume, the SNR threshold is chosen such that the false-alarm
probability at that threshold is small. In Refs.  \cite{Pekowsky:2012sr} and
\cite{Brown:2013hm} the same SNR threshold was used when estimating the
sensitive volume of dominant-mode templates and the potential sensitive volume
of sub-dominant mode templates.  Both studies acknowledged that adding
sub-dominant modes to templates can increase the probability of getting a false
alarm.  To keep the false-alarm probability fixed, the SNR threshold in a
search that uses sub-dominant mode templates must therefore increase. This
increase in threshold decreases the sensitive volume that can be obtained by
sub-dominant mode templates, making the case for adding sub-dominant modes to
templates weaker.

Due to these conflicting factors it is difficult to make a definitive statement
about the impact of sub-dominant modes on BBH searches from SNR considerations
alone. In this paper we resolve the uncertainty by finding both SNR and
$\chi^2$ between templates without sub-dominant modes and signals with
sub-dominant modes. Doing so, we are able to estimate the fraction of sensitive
volume that is lost by neglecting sub-dominant modes when re-weighted SNR is
used. We also simulate a search with sub-dominant modes: we estimate the
increase in threshold needed to keep the false-alarm probability fixed, thereby
allowing an accurate comparison of search volumes when sub-dominant modes are
included and excluded in templates. We consider non-spinning BBHs with
component-masses $3\,\Msun \leq m_1, m_2 \leq 200\,\Msun$ and with total mass
$M < 360\,\Msun$.  We therefore cover the entire range of ``stellar-mass" BBHs
that were searched for in LIGO and Virgo data in the past  ($m_1, m_2 \in [3,
97]\,\Msun;~M \leq 100\,\Msun$) \cite{\sFiveHighMass, \sSixHighMass}, and we
cover binaries that involve ``intermediate-mass black holes" which may form
from dynamical capture in globular clusters \cite{Miller:2004IMBH}. To generate
both dominant-mode and sub-dominant mode waveforms we use the EOB model calibrated
to numerical-relativity simulations, as obtained in Ref. \cite{Pan:2011gk}.

For simplicity, and as was done in Refs. \cite{Pekowsky:2012sr, Brown:2013hm},
we study the effect of sub-dominant modes using a single detector (real
searches use a network of detectors). We simulate an advanced LIGO detector by
generating stationary Gaussian noise colored by the zero-detuned, high-power
advanced LIGO design curve \cite{Shoemaker2009}. Due to the presence of
non-Gaussian transients there is currently no model for the noise distribution
of real detector data \cite{Aasi:2012wd, McIver:2012dq}. However, by injecting
signals into Gaussian noise we find the best sensitivity that can be obtained
by the search. We reason that if sub-dominant mode templates have worse
sensitivity than dominant-mode templates in Gaussian noise, those templates
will fare no better in real detector data.  Furthermore, current detection
pipelines are able to mitigate noise transients such that the sensitive volumes
of real detectors are within a factor of a few of what they would be if the
data were Gaussian \cite{ihopePaper:2012}. The sensitive volumes we find using
Gaussian noise is thus a good approximation of what they will be in real
detector data (assuming advanced detectors have similar data-quality
characteristics).

In this paper we are concerned primarily with the effects of sub-dominant modes
on our ability to \emph{detect} gravitational waves; we do not address the
effect on \emph{parameter estimation}. The goal of detection is to determine
whether or not a signal exists in some data, regardless of the parameters. When
doing parameter estimation, on the other hand, a signal is assumed to be in the
data; the goal is then to find the best fitting parameters.\footnote{In prior
searches, a detection pipeline was run on data first to identify times when
candidate signals exist; these times were then followed up by parameter
estimation pipelines \cite{Aasi:2013kqa}.} These differing goals put different
constraints on what to use as template waveforms. As we will see, including
sub-dominant modes in templates does not improve our ability to detect if the
sub-dominant modes do not increase the SNR of signals enough to offset
increases in false-alarm probability.  If a signal is assumed to be present,
however, then adding sub-dominant modes to template waveforms can only improve
parameter estimation if the resulting waveform is a better match to the signal.
Indeed, it has been shown \cite{McKechan:2011ps, Littenberg:2012uj} that
including sub-dominant modes reduces systematic bias when measuring the
parameters of signals.

The rest of this paper is divided as follows: Sec. \ref{sec:background}
provides background for the search methods and statistics we discuss.  In Sec.
\ref{ssec:DMsnr} we review how SNR is calculated using dominant-mode templates;
in Sec. \ref{ssec:chisq} we review the $\chi^2$ statistic and how it is used to
re-weight the SNR; in Sec. \ref{ssec:statistics} we discuss the statistics we
use in this paper to compare the sensitivity of searches; in Sec.
\ref{ssec:astrophysics} we provide a brief review of the astrophysics of BBHs
to motivate our choice of masses and rate priors. We show that the bank of
dominant-mode templates we use is effectual to dominant-mode signals across the
mass space we investigate in Sec.  \ref{ssec:effDMBank}.  In Sec.
\ref{ssec:HMeff} we find the sensitivity of this bank to signals with
sub-dominant modes to see if sub-dominant modes have an effect on predicted
sensitivity.  In Sec. \ref{sec:effcyOfHMbank} we estimate the sensitivity of a
simulated bank of sub-dominant mode templates using equations derived in
Appendices \ref{appdx:bankFAP} and \ref{appdx:maxHMsnr-analytic}. Finally, in
Sec. \ref{sec:conclusions} we review and discuss our results.

\section{Review of current searches and template modeling}
\label{sec:background}

The strain induced in a detector from a passing gravitational wave is
\cite{FinnChernoff:1993}:
\begin{align}
\label{eqn:strain}
h(t;\intP, \extP) &= \frac{1}{r} \big( F_\pls(\alpha, \delta, \psi) h_\pls(t-t_c; \theta, \phi, \phi_0, \intP) \nonumber \\
    &\qquad + F_\crs(\alpha, \delta, \psi) h_\crs(t-t_c; \theta, \phi, \phi_0, \intP) \big),
\end{align}
where:
\begin{flalign}
\label{eqn:hplus_hcross}
h_{(\pls,\crs)} &= (\Re,\Im) \sum_{lm} {}_{-2}Y_{l m}(\theta, \phi) A_{lm}(t-t_c; \intP) \nonumber \\
& \qquad \quad \times \exp\left[-i \Big(\Psi_{lm}(t-t_c; \intP) + m\phi_0\Big)\right]. &
\end{flalign}
Here, ${}_{-2}Y_{l m}(\theta, \phi)$ are the $-2$ spin-weighted spherical
harmonics, $r$ is the distance to the source from the detector, $\phi_0$ is the
initial phase of the binary, and $t_c$ is the coalescence time of the binary.
The angle $\theta$ is the angle between the orbital-angular momentum and the
line-of-sight to the detector (the \emph{inclination}); $\phi$ is the azimuthal
angle to the projection of the line-of-sight on to the orbital plane.  The
functions $F_\pls$ and $F_\crs$ project the gravitational wave from the
source's radiation frame into the detector's frame; they are functions of the
right ascension ($\alpha$), declination ($\delta$), and polarization ($\psi$)
of the source with respect to the detector
\cite{1987MNRAS.224..131S}.\footnote{Due to the motion of the Earth, $F_\pls$
and $F_\crs$ are also functions of time. Here we assume that the relative
displacement of the detector is small across the duration of the signal in the
detector's band.} Since we only consider non-spinning waveforms in this paper,
the \emph{intrinsic} parameters $\intP$ are the component masses $m_1$ and
$m_2$.  Neglecting spin also means that $\phi$ and $\phi_0$ are degenerate, and
we can set $\phi_0 = 0$. Together the parameters $\{t_c, r, \theta, \phi,
\alpha, \delta, \psi\}$ make up the \emph{extrinsic} parameters $\extP$. 

\subsection{SNR maximization for a dominant mode bank}
\label{ssec:DMsnr}

Given some detector data $s$ we wish to determine whether or not a
gravitational-wave signal $h$ is present in it. We do not know \emph{a priori} the
parameters of $h$. In order to maximize the probability of detection we must
therefore search over all possible intrinsic and extrinsic parameters that $h$
may have. To do this, we calculate the signal-to-noise (SNR) $\rho$ maximized
over $\intP$ and $\extP$:
\begin{equation}
\label{eqn:defSNR}
\rho = \max_{\intP, \extP} \frac{ \ip{h(t; \intP, \extP)}{s(t)} }{ \sqrt{\ip{h(t; \intP, \extP)}{h(t; \intP, \extP)}} }.
\end{equation}
The inner product $\ip{\cdot}{\cdot}$ is defined as \cite{Allen:2005fk}:
\begin{equation}
\label{eqn:defIP}
\ip{a}{b} = 4 \Re \int_{0}^{\infty} \frac{\tilde{a}^*(f) \tilde{b}(f)}{S_n(f)} \d f,
\end{equation}
where $S_n(f)$ is the one-sided power spectral density (PSD) of the noise. In this
paper we use the zero-detuned, high-power, advanced LIGO design curve
\cite{Shoemaker2009}. This PSD grows substantially at frequencies below $\sim
10\,$Hz due to seismic noise. As was done in Ref. \cite{Brown:2013hm}, we use a
lower frequency cutoff of $15\,$Hz for our matched filter. We terminate the
filter at a frequency larger than the ringdown frequency of the signal.

In principle we must maximize the SNR over 8 parameters for non-spinning
systems --- 2 intrinsic plus 6 extrinsic\footnote{We do not need to maximize
over $r$ since it cancels in the SNR.} --- but this number can be reduced. Let:  
\begin{align}
|F| &= \sqrt{F_\pls^2 + F_\crs^2}, \\
\kappa &= \arctan\left(\frac{F_\crs}{F_\pls}\right);
\end{align}
then:
\begin{equation}
h = \frac{1}{\mathcal{D}}\left(h_\pls \cos\kappa + h_\crs \sin\kappa\right),
\end{equation}
where $\effdist = |F|/r$ is the \emph{effective distance} \cite{Allen:2005fk}.
Since $\effdist$ cancels in the SNR, we need not maximize over it. We can thus
maximize over $\alpha,~\delta,$ and $\psi$ by simply maximizing $\kappa$.

The coalescence time is maximized over by evaluating the SNR at discrete time
intervals, selecting points where $\rho$ is at a maximum and exceeds some
threshold. These points are \emph{triggers}. In Ref. \cite{Allen:2005fk} it is
shown that $\rho(t)$ can be efficiently calculated by taking the inverse
Fourier transform of $\tilde{h}^*(f)\tilde{s}(f)/S_n(f)$. Since we are
interested in the effect of sub-dominant modes, which do not affect the
maximization over coalescence time, we will make this maximization implicit and
set $t_c = 0$ throughout the rest of this paper.

In order to maximize over the intrinsic parameters a template bank is used
\cite{Balasubramanian:1995bm, Owen:1995tm}. Templates are typically laid out
across the search parameter space such that no more than $3\%$ of the SNR is
lost due to the discreetness of the bank \cite{\sSixLowMass,\sSixHighMass}. The
SNR is maximized over the extrinsic parameters for each trigger; the template
with the largest SNR is then selected, thereby maximizing over the intrinsic
parameters.

With these considerations Eq. \eqref{eqn:defSNR} simplifies to:
\begin{equation}
\label{eqn:simplifiedSNR}
\rho = \max_{\theta, \phi, \kappa} \frac{ \ip{h_\pls(t; \theta, \phi)}{s(t)}\cos\kappa + \ip{h_\crs(t; \theta, \phi)}{s(t)}\sin\kappa}{\sqrt{\ip{h(t; \kappa, \theta, \phi)}{h(t; \kappa, \theta, \phi)}}}.
\end{equation}
If the templates contain sub-dominant modes, then the number of parameters that
need to be maximized over can be reduced no further. In Appendix
\ref{appdx:maxHMsnr-analytic} we perform the maximization over $\kappa$
analytically when sub-dominant modes are included (the maximization over
$\theta$ and $\phi$ must be done numerically). This is used to model a
hypothetical search using sub-dominant modes, which is discussed in Sec.
\ref{sec:effcyOfHMbank}. In current searches, template waveforms are generated
using only the dominant, $l = |m| = 2$, mode. In that case $\theta,~\phi$,
and $\kappa$ are all degenerate with each other and the maximization reduces to
a single parameter. This can be performed analytically, yielding
\cite{Sathyaprakash:1991mt}:
\begin{equation}
\label{eqn:dmSNR}
\rho = \sqrt{\frac{\ip{h_\pls(t)}{s(t)}^2 + \ip{h_\crs(t)}{s(t)}^2}{\ip{h}{h}}}.
\end{equation}
In stationary Gaussian noise containing no signal, $\rho$ is $\chi$ distributed
(or equivalently, $\rho^2$ is $\chi^2$ distributed) with two degrees of freedom.

\subsection{The $\chi^2$ test and re-weighted SNR}
\label{ssec:chisq}

Real data from the LIGO and Virgo detectors contain a number of non-Gaussian
transients (\emph{glitches}) \cite{Aasi:2012wd, McIver:2012dq}. To mitigate the
effect of these glitches the $\chi^2$ statistic is calculated to better
separate noise from potential signals. This is calculated as follows: split the
matched filter into $p$ frequency bins such that the template has equal amounts
of power in each bin, and let $\rho_i$ be the SNR of a trigger in the $i^\th$
bin. If the signal matches the template then $\rho_i \approx \rho/p$. We can
therefore quantify how well a signal matches a template by defining
\cite{Allen:2004}:
\begin{equation}
\label{eqn:defChisq}
\chi^2 = p \sum_{i=1}^{p} \left|\rho_i - \frac{\rho}{p}\right|^2.
\end{equation}
If the template matches the signal exactly then the $\chi^2$ statistic will be
$\chi^2$ distributed with $2p-2$ degrees of freedom (hence the name). If there
is any mismatch between the signal and the template, the mean $\chi^2$ is
\cite{Allen:2004}:
\begin{equation}
\label{eqn:chisqMismatch}
\left<\chi^2\right> = 2p - 2 + \mu^2\rho^2,
\end{equation}
where $\mu$ is a measure of the mismatch. Thus the larger the mismatch between
the signal and the template, the larger the $\chi^2$ statistic. Since the
increase in $\chi^2$ is proportional to $\rho$, even ``loud" (large SNR)
glitches are mitigated by $\chi^2$ as they will have high mismatch to the
template.

In BBH searches $\chi^2$ is used to re-weight $\rho$, obtaining the \emph{re-weighted
SNR} $\rhonew$.\footnote{A cut on large $\chi^2$ values is also employed, but this cut is
chosen conservatively in order to not remove any signals \cite{ihopePaper:2012}.
As such, it is mostly used to reduce the number of triggers for computational
purposes. We therefore do not consider it here.} As it is based on
characteristics of the data and the templates, the exact form of the weighting
evolved throughout initial LIGO. Generally, the weighting is such that $\rho$
is reduced for triggers with high $\chi^2$. In this study we use the
re-weighted SNR that was used in the most recent searches
\cite{\sSixLowMass,\sSixHighMass}, which is defined as:\footnote{In Ref.
\cite{\sSixHighMass}, a different weight than that given in Eq.
\eqref{eqn:newsnr} was used for triggers whose templates had duration $< 0.2\,$s.
All of the templates in our study have durations $> 1\,$s, however, due to
the better low-frequency performance of the advanced LIGO noise curve we use.
We therefore use Eq. \eqref{eqn:newsnr} for all triggers.} 
\begin{equation}
\label{eqn:newsnr}
\rhonew = \left\{
\begin{array}{cl}
\rho & \text{for $\chi_r^2 \leq 1$}, \\
\rho
\left[\frac{1}{2} \left(1 +
\left(\chi_r^2\right)^{\!3}\right)\right]^{-1/6}
& \text{for $\chi_r^2 > 1$}.
\end{array} 
\right.
\end{equation}
where $\chi^2_r$ is the \emph{reduced $\chi^2$}, which is $\chi^2$ divided by
the number of degrees of freedom.

Re-weighting the SNR using $\chi^2$ is crucial to BBH searches. An SNR of 8 is
typically used when predicting detection rates \cite{\RatesPaper}, but it would
not be possible to detect at this $\rho$ without some type of $\chi^2$
re-weighting of glitches \cite{ihopePaper:2012}. However, if signals also
significantly mismatch the templates then the use of $\rhonew$ can adversely
affect the efficiency to these signals. The effect on efficiency from the
mismatch between signals and templates therefore cannot be determined from the
loss in SNR alone. For this reason, in the following sections we investigate
the effect on $\chi^2$ due to the mismatch between dominant-mode templates and
signals with sub-dominant modes, and we calculate efficiency using $\rhonew$.

\subsection{Quantifying the ability of a search to make detections}
\label{ssec:statistics}

The purpose of our study is to investigate whether adding sub-dominant modes to
templates will improve our ability to detect gravitational waves from
non-spinning BBHs. Ultimately, we want to know what template bank maximizes the
number of detections made per unit time. To do so we need to quantify the
ability of a bank with dominant-mode templates and a bank of templates with
sub-dominant modes to recover waveforms with sub-dominant modes. For this
purpose we calculate \emph{effectualness}; we also calculate \emph{efficiency},
from which we find \emph{sensitive volume} and \emph{relative gain}.

\subsubsection{Effectualness}
\label{sssec:effectualness}

Effectualness\footnote{Effectualness has also been referred to as ``fitting
factor" \cite{Apostolatos:1995pj}.} is a statistic that is commonly used to
quantify how well one family of waveforms recovers another. Given a simulated
signal $h^\dagger$ with parameters $\allP = (\intP, \extP)$ and a bank of
templates $\{h(\allP')\}$, effectualness is defined as
\cite{Apostolatos:1995pj,Damour:1997ub}:
\begin{equation}
\label{eqn:defEff}
\eff{a}{b}(\allP) = \max_{\allP'} \frac{\ip{h_a(\allP')}{h_b^\dagger(\allP)}}{\sqrt{\ip{h_a(\allP')}{h_a(\allP')}\ip{h_b^\dagger(\allP)}{h_b^\dagger(\allP)}}}.
\end{equation}
Here and elsewhere we adopt subscripts on $\eff{}{}$ to indicate whether or not
the signal and templates have sub-dominant modes; the first index indicates the
templates, the second the signal. We will use $\sd$ to indicate a waveform that
has sub-dominant modes and $\dm$ to indicate a waveform without sub-dominant
modes.  For example, $\eff{\dm}{\sd}(\allP)$ is the effectualness of a bank of
templates without sub-dominant modes to a signal with sub-dominant modes and
parameters $\allP$.

If a signal $h^\dagger$ is in stationary Gaussian noise with zero-mean, the
expectation value of the overlap with a template $h$ is $\ip{h}{s = h^\dagger +
n} = \ip{h}{h^\dagger}$.  The maximum overlap occurs when $h$ and $h^\dagger$
have the same parameters and come from the same waveform model.  Thus, via Eq.
\eqref{eqn:defSNR}, the expectation value of the maximum recoverable SNR of a
signal $h^\dagger$ is:
\begin{equation}
\label{eqn:maxSNR}
\max \left<\rho\right> = \sqrt{\ip{h^\dagger}{h^\dagger}}.
\end{equation}
Effectualness therefore gives the fraction of available SNR in a signal $h^\dagger_b$
that is recovered by the template $h_a$:
\begin{equation}
\label{eqn:eff2rho}
\eff{a}{b}(\allP) = \frac{\left<\rho_{ab}(\allP)\right>}{\max \left<\rho_{bb}(\allP)\right>}.
\end{equation}
Here, $\left<\rho_{ab}(\allP)\right>$ indicates the expectation value of the
SNR of signal $h_b^\dagger(\allP)$ using templates
$\left\{h_a(\allP')\right\}$.\footnote{Note that $\max
\left<\rho_{bb}(\allP)\right>$ is not the same as
$\left<\rho_{bb}(\allP)\right>$.  This is because
$\left<\rho_{bb}(\allP)\right>$ indicates the SNR of a bank of templates
$\{h_{b}(\allP')\}$ to a signal $h^\dagger_{b}(\allP)$. In fact, while the
waveform models may be the same, the template and signal parameters may not due
to the discreetness of the bank.} The smaller the effectualness, the less
SNR recovered by the template and the closer signals need
to be in order to detect them.  Effectualness is thus an estimate of how
sensitive a bank of templates will be to a particular set of signals.

However, effectualness is not sufficient for comparing the sensitivity of a set
of templates to signals created with different waveform models. We see from Eq.
\eqref{eqn:eff2rho} that a drop in effectualness can result from a decrease in
the overlap between two waveforms or from an increase in the maximum
recoverable SNR. This ambiguity has particular relevance for sub-dominant
modes. The predicted detection rates of advanced LIGO are made assuming that
both signals and templates have the dominant mode only \cite{\RatesPaper}. Real
signals will have sub-dominant modes. If $\eff{\dm}{\sd} < \eff{\dm}{\dm}$ for
some signals, it is not clear from effectualness alone whether the loss is due
to $\left<\rho_{\dm \sd}\right> < \left<\rho_{\dm\dm}\right>$ or if it is
because $\max \left<\rho_{\sd \sd}\right> > \max \left<\rho_{\dm \dm}\right>$.
If entirely the first case, detection rates will be less than predicted.  If
entirely the second case, the sensitivity of aLIGO will be the same as
predicted (perhaps better, since the additional power in the sub-dominant modes
may give larger SNR than expected). The drop in effectualness in this case only
indicates that the sensitivity could be better than predicted if sub-dominant
modes were added to templates, but it will not be worse.

It is also not clear what affect a decrease in effectualness has on re-weighted
SNR. Equation \eqref{eqn:chisqMismatch} indicates that larger mismatch between
signal and template results in larger $\chi^2$; this increase results in lower
$\rhonew$. Since re-weighted SNR is used as the ranking statistic in real
searches, the decrease in effectualness (increase in mismatch) may result in a
further reduction in sensitivity via $\chi^2$. 

Finally, effectualness implicitly assumes that SNR is the optimal statistic
(in the Neyman-Pearson sense) to detect a gravitational wave in Gaussian noise.
While this has been shown to be true when the intrinsic parameters of the
signal are known \cite{Sathyaprakash:1991mt}, it is not necessarily true if the
intrinsic parameters are unknown, as is the case in real BBH searches.  Finding
the optimal statistic in this case is difficult to do, as the waveforms have a
non-trivial dependence on the intrinsic parameters.  Moreover, we would need to
know the distribution of the sources' intrinsic parameters, which in our case
are the masses of BBHs.  As we will see in Sec.~\ref{ssec:astrophysics} the
mass distribution of BBHs is highly uncertain. Even if we could find the
optimal statistic for an assumed distribution, the statistic may not be optimal
for the real astrophysical distribution.

In practice, we get around these difficulties by simply assuming that SNR
maximized over the template bank is a good approximation to the optimal
statistic.  In doing so, we assume that each template is equally likely to
detect a signal, which implicitly assumes a particular astrophysical
distribution. Effectualness then gives us a measure of the performance of the
bank assuming that distribution is correct. However, if the implicit
distribution is not correct, then SNR maximized over the bank is not a good
approximation of the optimal statistic, thereby making effectualness a poor
metric for search sensitivity.

For these reasons the effect of sub-dominant modes on the sensitivity of a
search cannot be ascertained by effectualness alone. A more informative metric
is the sensitive volume, which is found from efficiency.

\subsubsection{Efficiency, sensitive volume, and relative gain}
\label{sssec:eff_sV_rG}

Given a set of $N$ simulated signals with intrinsic and extrinsic parameters
$\allP$ and generated from waveform-model $b$, $\{h^\dagger_b(\allP)\}$, the
efficiency of a bank of templates generated from waveform-model $a$ is the
fraction of signals found with $\rhonew$ larger than some threshold value at a
distance $r$ \cite{loudestGWDAW03}; i.e.,
\begin{equation}
\label{eqn:effcyDef}
\effcy{a}{b}(r, \intP) = \frac{n_{a b}(r, \intP)}{N_{b}(r, \intP)}.
\end{equation}

As with effectualness subscripts indicate the type of waveform used for the
templates and signals; e.g., $\effcy{\dm}{\sd}$ is the efficiency of a dominant
mode bank to sub-dominant mode signals. Note that, aside from $r$,
$\effcy{a}{b}$ is a function of the intrinsic parameters only ($\intP$).  Since
the universe is isotropic at the distances we are considering, the rest of the
extrinsic parameters are accounted for by uniformly distributing signals in
$\{\theta, \phi,\phi_0, \alpha, \delta, \psi\}$.

When determining efficiency in real searches the threshold $\rhonew$ is
determined by the loudest event in the data \cite{\sSixLowMass,\sSixHighMass}.
In this study we use $\rhonew = 8$ as the threshold. SNR ($\rho$) equal to 8 is
commonly used as a threshold for predicting detection rates \cite{\RatesPaper}.
For well-matched signals, $\rhonew \approx \rho$ when $\rho = 8$; using
$\rhonew = 8$ as a threshold should therefore give roughly the same results as
predictions if mismatches between signals and templates are not too large. 

Integrating the efficiency times the astrophysical rate of BBHs $R(V, \intP)$
over volume and $\intP$ gives the expected rate of GW detections per unit time:
\begin{equation}
\mathcal{R}_{\mathrm{detect}} = \int \epsilon(r, \intP)R(V, \intP)\d V \d\intP.
\end{equation}
Assuming $R(V, \intP)$ is constant over volumes for which $\epsilon \neq 0$ and
uniform in a small region in parameter space $\intP + \Delta \intP$, we define
the sensitive volume $V_{ab}$ as:
\begin{equation}
\label{eqn:sensitiveV}
V_{ab}\left(\intP + \Delta \intP\right) = 4\pi \int_{0}^{\infty} \effcy{a}{b}\left(r, \intP + \Delta \intP\right) r^2 \d r.
\end{equation}
With these assumptions, $V_{ab}$ is proportional to the average rate of
detections in $\intP + \Delta \intP$.

We assume that all of the error in the sensitive volume is due to error in the
measurement of the efficiency from statistical fluctuations in the noise. This
error is derived from the range of efficiency values for which the number of
found injections varies no more than one standard deviation \cite{Agresti:1998}:
\begin{equation*}
(n - \left<n\right>)^2 \leq \left<n^2\right> - \left<n\right>^2.
\end{equation*}
The mean and variance are given by the Binomial distribution:
\begin{align*}
\left<n\right> &= N\epsilon, \\
\left<n^2\right> &= N\epsilon(1 - \epsilon).
\end{align*}
Using these values and solving for $\epsilon$ yields \cite{Agresti:1998}:
\begin{equation}
\label{eqn:epsilon_err}
 \pm\delta\effcy{i}{} = \frac{N_{i} (2n_{i} + 1) \pm \sqrt{4N_{i}n_{i}(N_{i}-n_{i}) + N^2_{i}}}{2N_{i}(N_{i}+1)},
\end{equation}
where the index $i$ indicates the efficiency and number of injections in the
$i^\th$ distance bin (dependence on waveforms $a,b$ and parameters $\intP$ made
implicit). This error is then propagated to the volume via Eq.
\eqref{eqn:sensitiveV}. Note that this results in a conservative estimate of
$\delta V$. In our study we average over several realizations of noise when
finding $\rhonew$; thus actual statistical fluctuations in $\effcy{a}{b}$ are
less than what is assumed in Eq. \eqref{eqn:epsilon_err}.

To compare the sensitivity of two template banks to a set of simulated signals
we define the relative gain $\gain{cd}{ab}$:
\begin{equation}
\label{eqn:gainDef}
\gain{cd}{ab}(\intP + \Delta\intP) = \frac{V_{cd}(\intP + \Delta\intP)}{V_{ab}(\intP + \Delta\intP)}.
\end{equation}
The upper (lower) indices denote the waveform models used for templates and
signals in the numerator (denominator). Since sensitive volume is proportional
to the detection rate, the relative gain gives the average number of detections
per unit time the $cd$ search will make relative to the $ab$ search.

\subsection{Astrophysical priors on the distribution of BBH masses}
\label{ssec:astrophysics}

When comparing the sensitivity of template banks it can happen that one bank
has better sensitivity in one area of parameter space, but worse sensitivity in
another. We will see this in Sec. \ref{sec:effcyOfHMbank}; templates with
sub-dominant modes have better sensitivity at $M > 100\,\Msun$ and $q \gtrsim 10$,
but worse sensitivity for lower mass and more equal-mass systems. This raises
the question of whether the gain in sensitivity in one area of parameter space
is large enough to offset the loss in another.  In this case it is useful to
try to find the \emph{net} relative gain, which is:
\begin{equation}
\label{eqn:netGainDef}
\netgain{cd}{ab} = \frac{\int R(\intP) V_{cd}(\intP) \d \intP}{\int R(\intP) V_{ab}(\intP) \d \intP}.
\end{equation}
Doing so gives the relative number of detections made over the entire parameter
space; if the net gain is $> 1$ the gain in sensitivity in one area of parameter
space is worth the loss in another.

Calculating net gain requires knowledge of the rate of BBH coalescence,
$R(\intP)$. Unfortunately, this is highly uncertain. For a BBH with $m_1 = m_2
= 10\,\Msun$, the coalescence rate has been estimated to be anywhere from
$O(10^{-4})$ to $O(10^{-1})\,\mathrm{Mpc}^{-3}\mathrm{Myr}^{-1}$, leading to
predicted detection rates between $0.4$ and $10^3$ per year in advanced LIGO
\cite{\RatesPaper}. We do not need to know the magnitude of the BBH coalescence
rate in order to calculate the net gain --- we only need the relative distribution
in mass --- but even this is largely uncertain.

Binary black holes with masses detectable by LIGO are thought to be formed in
one of two ways: via the two stars in an isolated binary each collapsing into
black holes (\emph{field binaries}), or by a black hole capturing another black
hole in a dense stellar region such as globular clusters
\cite{LVCwhitepaper:2012}. No BBHs have been directly observed; their existence
is predicted from population synthesis models, from the predicted evolution of
known X-ray binaries, and from considerations of the dynamics of black holes in
dense stellar clusters \cite{\RatesPaper}.

The most massive black holes formed from stellar collapse known are in the
X-ray binaries IC 10 X-1 and NGC 300 X-1 \cite{Silverman:2008, Crowther:2010}.
These black holes have been estimated to have masses between $20$ and
$35\,\Msun$, but population synthesis models predict that black holes formed
from isolated field stars may have masses as large as $80\,\Msun$ in
low-metallicity environments \cite{Belczynski:2009}. The mass distribution of
black holes in field binaries are more difficult to predict, however, as the
proximity of the two progenitor stars to each other add several complications
\cite{Dominik:2012}.  Population synthesis models give varying results
depending on the values assumed for input parameters, but models generally
suggest field binaries have $M \lesssim 100\,\Msun$ \cite{Dominik:2013}, with
the peak of the distribution occurring around $20\,\Msun$ (cf. Figs. 8 and 9 of
Ref.  \cite{Dominik:2013}).  The distribution in mass ratio is also uncertain:
some models predict equal-mass systems are more likely, while others predict a
roughly uniform distribution between $q = 1$ and $4$ (cf. Fig. 9 of Ref.
\cite{Dominik:2012}).

Black holes formed from dynamical capture in globular clusters may have masses
between $10^2$ and $10^4\,\Msun$ \cite{Miller:2004IMBH}. The existence of these
intermediate-mass black holes (IMBHs) is more speculative, but are supported by
observations of ultra luminous X-ray sources \cite{Miller:2009}. If IMBHs do
exist they may form binaries with other IMBHs or with stellar-mass black holes,
thus forming BBHs with larger total masses and mass ratios than possible in
field binaries \cite{\RatesPaper}. The merger rate between an IMBH with mass
between $50$ and $350\,\Msun$ and a stellar-mass black hole with mass $m$ has
been estimated to be $\sim (0.02/m)\, \mathrm{Mpc}^{-3} \mathrm{Myr}^{-1}$
\cite{Brown:2007}. This is an optimistic estimate \cite{\RatesPaper}, but if
correct, the rate of these \emph{intermediate-mass-ratio inspirals} (IMRIs)
could be on the same order of magnitude as BBHs formed from field binaries.

In this paper we consider BBHs with $3 \leq m_1, m_2 \leq 200\,\Msun$ and
$M\leq 360\,\Msun$. We therefore cover the entire predicted mass range of field
binaries and the lower end of IMBH/IMBH and IMRI binaries. (Our choice of
masses is based on the effectualness of the template placement algorithm we
use; see Sec. \ref{sec:HMbank} for details.) Due to the large uncertainty in
mass distribution across this range, we will simply assume a uniform rate in
$m_1$ and $m_2$ to calculate net gain. This choice of astrophysical prior
weights equal-mass systems as being more likely to occur than asymmetric-mass
systems.  Since this may bias our results (sub-dominant modes are more
significant in higher mass-ratio systems) we will also consider a rate uniform
in $M$ and $q$.  We will find that BBHs with $M > 100\,\Msun$ are the dominant
contribution to the net gain using these rate priors, as the sensitive volumes
of these larger mass systems can be one to two orders of magnitude larger than
lower-mass BBHs.  Since the existence of IMBHs is more speculative, we
additionally report net gains when only stellar-mass BBHs ($M < 100\,\Msun$)
are considered.

The astrophysical rates we use are only meant to be rough approximations for
comparisons between template banks; we do not attempt to calculate detection
rates. We do, however, report sensitive volumes across the mass space [see Fig.
\ref{fig:sensitiveVol}].  If a particular astrophysical prior is chosen, our
results can be used to estimate the total number of non-spinning BBH detections
that will be made in advanced LIGO.

\section{Impact of sub-dominant modes on a dominant mode template bank}
\label{sec:HMbank}

Past BBH searches have used waveforms without sub-dominant modes as templates
\cite{\LigoSearchPapers}.  Predicted advanced LIGO detection rates are also
based on considerations of the dominant mode only \cite{\RatesPaper}. This
assumes that the effect of neglecting sub-dominant modes is small. To test the
validity of this assumption we compare the sensitivity of a template bank
without sub-dominant modes to simulated signals without sub-dominant modes and to
signals with sub-dominant modes.

We generate one million simulated signals (\emph{injections}) uniformly
distributed in component mass, inclination, sky location, distance, and
polarization. Component masses are between $3$ and $200\,\Msun$, resulting in
total masses ($M$) between $6$ and $400\,\Msun$ and mass ratios ($q$) between
$1$ and $\sim 66$. Distance limits are chosen such that an optimally-oriented
signal with the same masses as a given injection would produce an SNR between 6
and 100. For each injection we generate a version without sub-dominant
modes and a version with sub-dominant modes, both at the same physical
distance, using the EOB waveform calibrated to numerical relativity that is
described in Ref.  \cite{Pan:2011gk}.\footnote{Specifically, we use the EOBNRv2
code in the LSC Algorithm Library (LAL) \cite{LAL} to generate dominant-mode
waveforms. For sub-dominant mode waveforms we use the EOBNRv2HM code in LAL.}
Template waveforms are the dominant-mode version of this waveform.

To calculate the re-weighted SNR of each injection we generate 16 realizations of
Gaussian noise colored by the zero-detuned high power noise curve. We use a
different set of 16 realizations for each injection. In each realization we
calculate the maximum SNR before the injection is added to the noise and after.
If the SNR with the injection in the noise is larger than the SNR from noise
alone, we record the SNR and calculate $\chi^2$ and re-weighted SNR. We then
find the mean $\rho$, $\chi^2$, and $\rhonew$ to get a measure of their
expectation values. Only realizations that produced an SNR louder than noise
are included in the average.

Due to computational constraints we are not able to calculate re-weighted SNR
for every template for every injection. Instead, we use the best matching
template from the effectualness study to calculate $\rhonew$. This is
equivalent to maximizing on SNR, then finding re-weighted SNR, which is what current
searches do \cite{ihopePaper:2012}. We note that maximizing on $\rhonew$ might
give better sensitivity, but since this is not what is currently done we do not
investigate this idea further.

To calculate the sensitive volume we create 16 bins uniformly distributed in
$M$ and 10 bins uniformly distributed in $\log q$, keeping bins which have
at least $1000$ injections in them. Within each mass bin we create 20 bins
uniform in log distance. We find the efficiency in each distance bin, then
numerically integrate over the distance bins to get the sensitive volume. The
distance bins span the smallest injected distance to the largest within the
given mass bin. For distances smaller then the closest injection we assume
$\epsilon = 1$; for distances larger than the furthest injection we assume
$\epsilon = 0$.

\subsection{Effectualness of the dominant-mode bank}
\label{ssec:effDMBank}

Past BBH searches \cite{\sSixLowMass,\sSixHighMass} have constructed template
banks by placing templates in parameter space such that no more than $3\,\%$ of
SNR is lost due to the discreetness of the bank. In other words, templates are
placed such that the \emph{minimal match}, which is the minimum of the
effectualness of the bank, is $\geq 97\%$.  Templates are placed in a hexagonal
grid \cite{hexabank} using a metric calculated from the mismatch between a
waveform with intrinsic parameters $\intP$ and a waveform with parameters
$\intP + \delta \intP$ \cite{Balasubramanian:1995bm, Owen:1995tm}.  This metric
is found analytically using inspiral waveforms expanded to 1.5 post-Newtonian
order (PN) in phase \cite{Owen:1998dk}.\footnote{In Ref.~\cite{Owen:1998dk},
the metric was calculated to \emph{second} post-Newtonian order. However, a bug
was recently discovered in the template-placement code in LAL (on which this
study depends) that caused the metric calculation to be truncated at $1.5\,$PN.
Since this bug was discovered after we had finished all calculations, we simply
use the $1.5\,$PN metric.} As this metric is based solely on the inspiral part
of the waveform, it will fail to properly place templates at high mass, where
the waveforms are dominated by merger and ringdown \cite{BBCCS:2006}. More
sophisticated template placement methods have recently been devised: a new
metric has been calculated using the $3.5\,$PN approximant \cite{Keppel:2013tm}
and a ``stochastic" placement method has been developed that uses no metric
\cite{2012arXiv1210.6666A}. In this study we are only interested in the
relative difference in sensitivity when a template bank neglects sub-dominant
modes and not on the most efficient placement algorithm.  We therefore use the
same $1.5\,$PN placement algorithm that has been used in prior
searches,\footnote{We emphasize that while the templates are \emph{placed}
using the $1.5\,$PN approximant, the template waveforms are EOBNRv2.} putting a
cut on total mass where the metric fails to maintain the desired $97\,\%$
minimal match.

\begin{figure}
\includegraphics[width=\columnwidth]{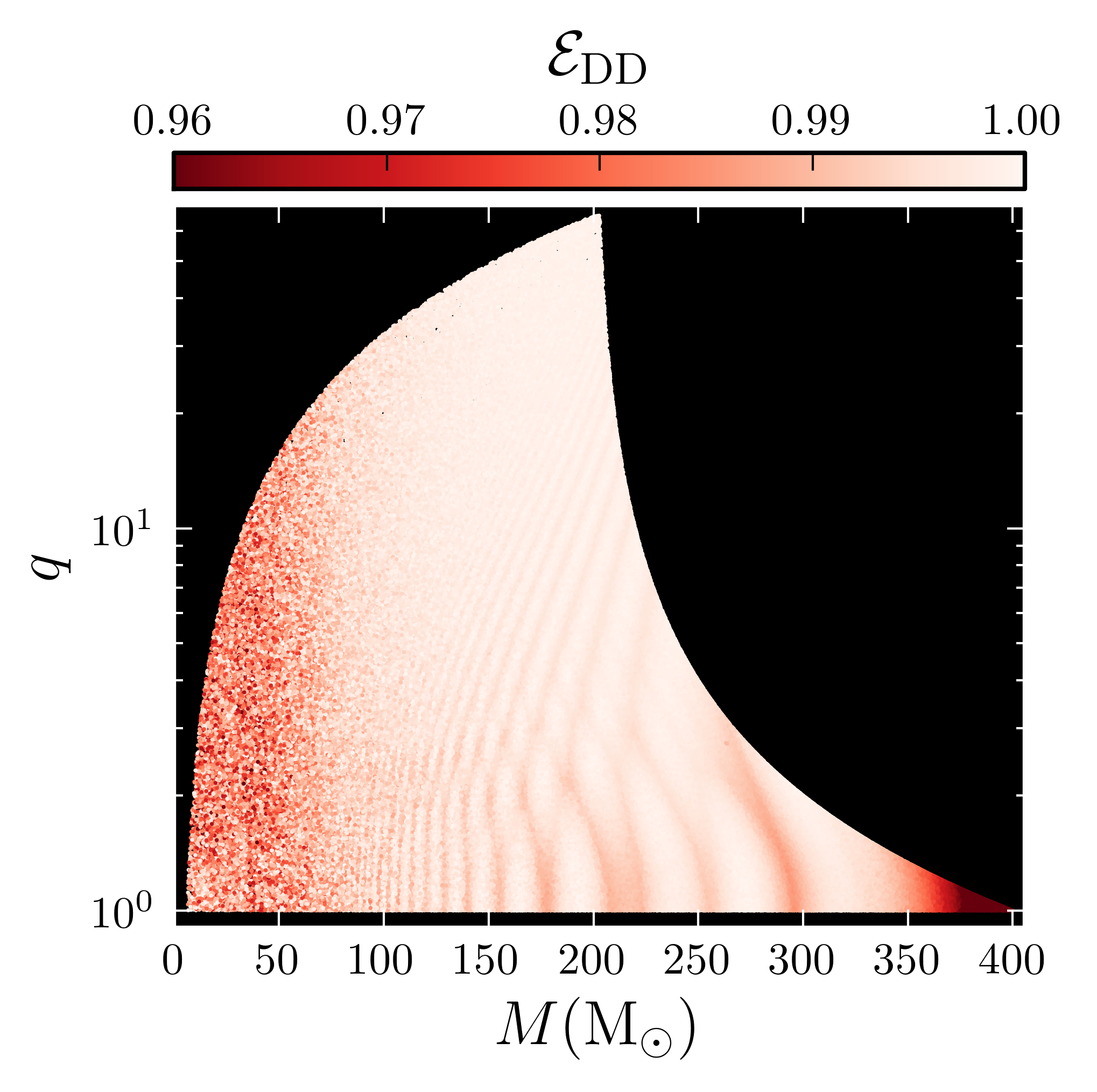}

\caption{Effectualness of a template bank placed using the $1.5\,$PN metric and using
waveforms without sub-dominant modes (EOBNRv2) to injections without
sub-dominant modes (also EOBNRv2), $\eff{\dm}{\dm}$. The desired minimal match
of the bank ($= \min\left\{\eff{}{}\right\}$) is 0.97.}

\label{fig:effbank-DM}
\end{figure}

To determine where the $1.5\,$PN metric fails to maintain a $97\,\%$ minimal
match we generate a template bank to cover component masses between $3$ and
$200\,\Msun$ ($6 \leq M/\Msun \leq 400$). Using the aLIGO zero-detuned high
power PSD, this results in a bank of $19\,800$ templates.  We calculate the
effectualness of this bank to the injections without sub-dominant modes,
$\eff{\dm}{\dm}$; the results are shown in Fig.  \ref{fig:effbank-DM}.  The
effectualness appears to be $>0.97$ for $M \lesssim 360\,\Msun$, but is lower
than that for larger $M$.  Indeed, we find that less than $0.3\,\%$ of
injections with $M<360\,\Msun$ have an effectualness $< 0.97$ [see Fig.
\ref{fig:effhist-HM}], but $\sim 70\,\%$ of injections have an effectualness $<
0.97$ for total masses larger than $360\,\Msun$. We therefore place a cutoff of
$M<360\,\Msun$ when reporting effectualness and efficiency.

\subsection{Efficiency to injections with sub-dominant modes}
\label{ssec:HMeff}

\begin{figure}
\includegraphics[width=\columnwidth]{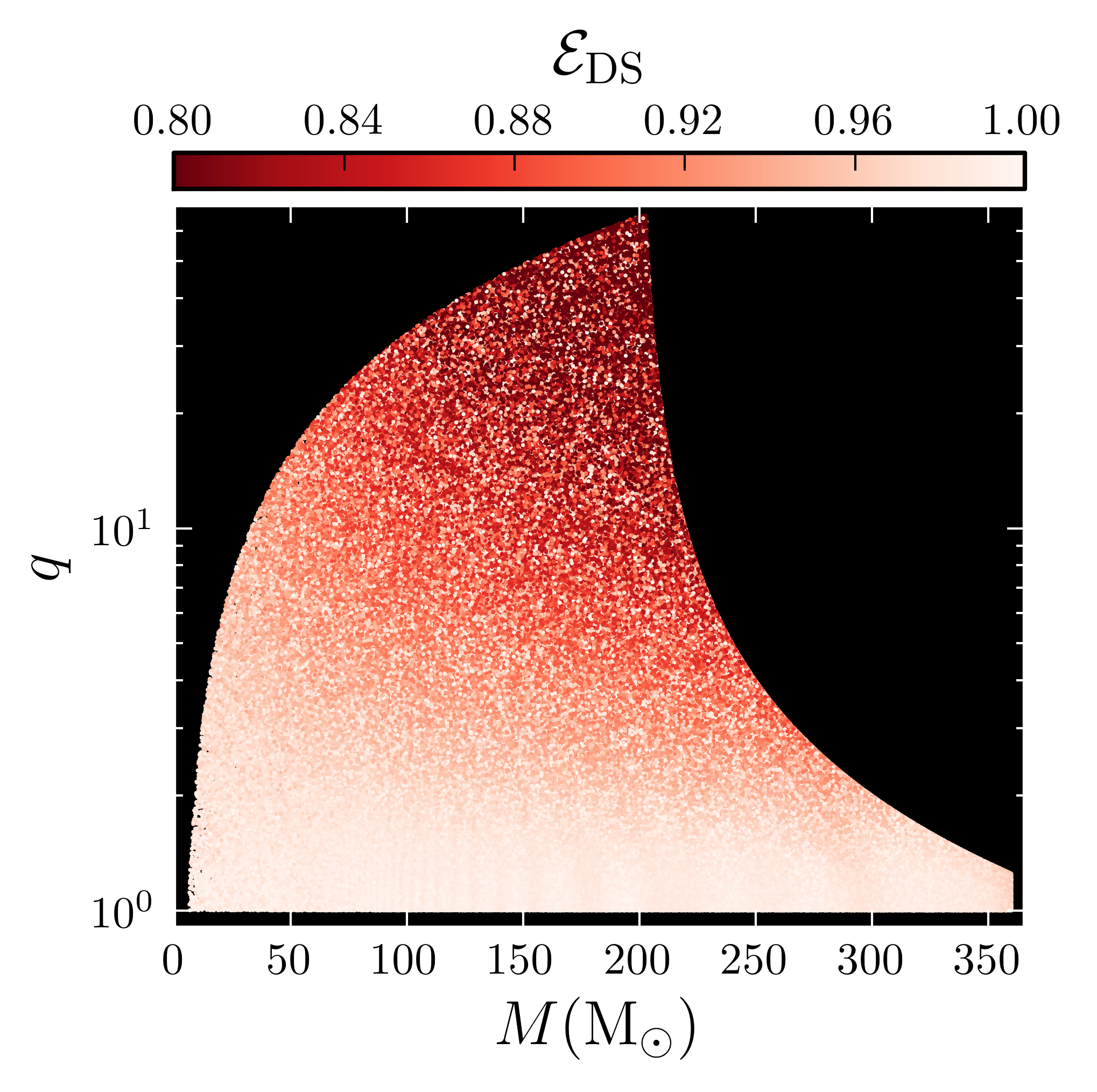}

\caption{Effectualness of the $1.5\,$PN dominant-mode template bank to
injections with sub-dominant modes, $\eff{\dm}{\sd}$. The desired minimal match
($= \min\left\{\eff{}{}\right\}$) of the bank was 0.97, but we see that
the effectualness drops below this as we go to higher mass ratio.}

\label{fig:HMeff}
\end{figure}

\begin{figure}
\includegraphics[width=\columnwidth]{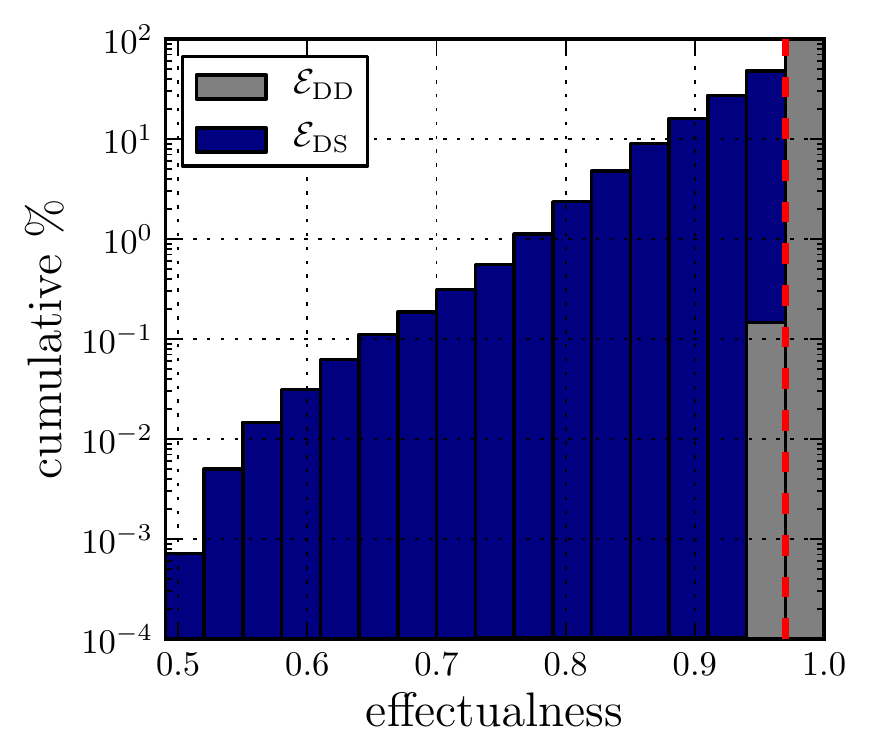}

\caption{Cumulative histogram of the effectualness of a template bank without
sub-dominant modes to injections without sub-dominant modes $\eff{\dm}{\dm}$
(gray) and injections with sub-dominant modes $\eff{\dm}{\sd}$ (blue). Each bin
gives the fraction of injections with an effectualness less than the right edge
of the bin. The red dashed line indicates the target minimal match for the bank
($0.97$). All injections have total mass $< 360\,\Msun$.}

\label{fig:effhist-HM}
\end{figure}

Figures \ref{fig:HMeff} and \ref{fig:effhist-HM} show the effectualness of the
dominant-mode bank when sub-dominant modes are added to the injections,
$\eff{\dm}{\sd}$. We find that $\eff{\dm}{\sd}$ is substantially lower than
$\eff{\dm}{\dm}$ for many injections: $\sim 50\,\%$ of sub-dominant injections
have an effectualness $<0.97$ for $M < 360\,\Msun$. In Fig. \ref{fig:HMeff}
we see that the effectualness decreases with increasing mass ratio. This is
expected as the amount of power in the higher modes will increase relative to
the dominant mode as the binary becomes more asymmetric. 

Given that $\eff{\dm}{\sd} < \eff{\dm}{\dm}$ we wish to know if the predicted
detection rates for advanced LIGO will be worse than expected. As discussed in
Sec. \ref{ssec:statistics} we cannot tell this from effectualness alone.
Instead, we compute $\gain{\dm\sd}{\dm\dm}$, which is the sensitive volume of
injections with sub-dominant modes ($\vol{\dm}{\sd}$) relative to the sensitive
volume of injections without sub-dominant modes ($\vol{\dm}{\dm}$) when both
are recovered by the dominant-mode template bank. If $\gain{\dm\sd}{\dm\dm} <
1$ it means that the sensitivity to real signals (which have sub-dominant
modes) will be worse than predicted.  If $\gain{\dm\sd}{\dm\dm} > 1$ it means
that the sensitivity to real signals will be better than predicted; if
$\gain{\dm\sd}{\dm\dm} = 1$ it means there is no difference.

\begin{figure*}
\includegraphics[width=2\columnwidth]{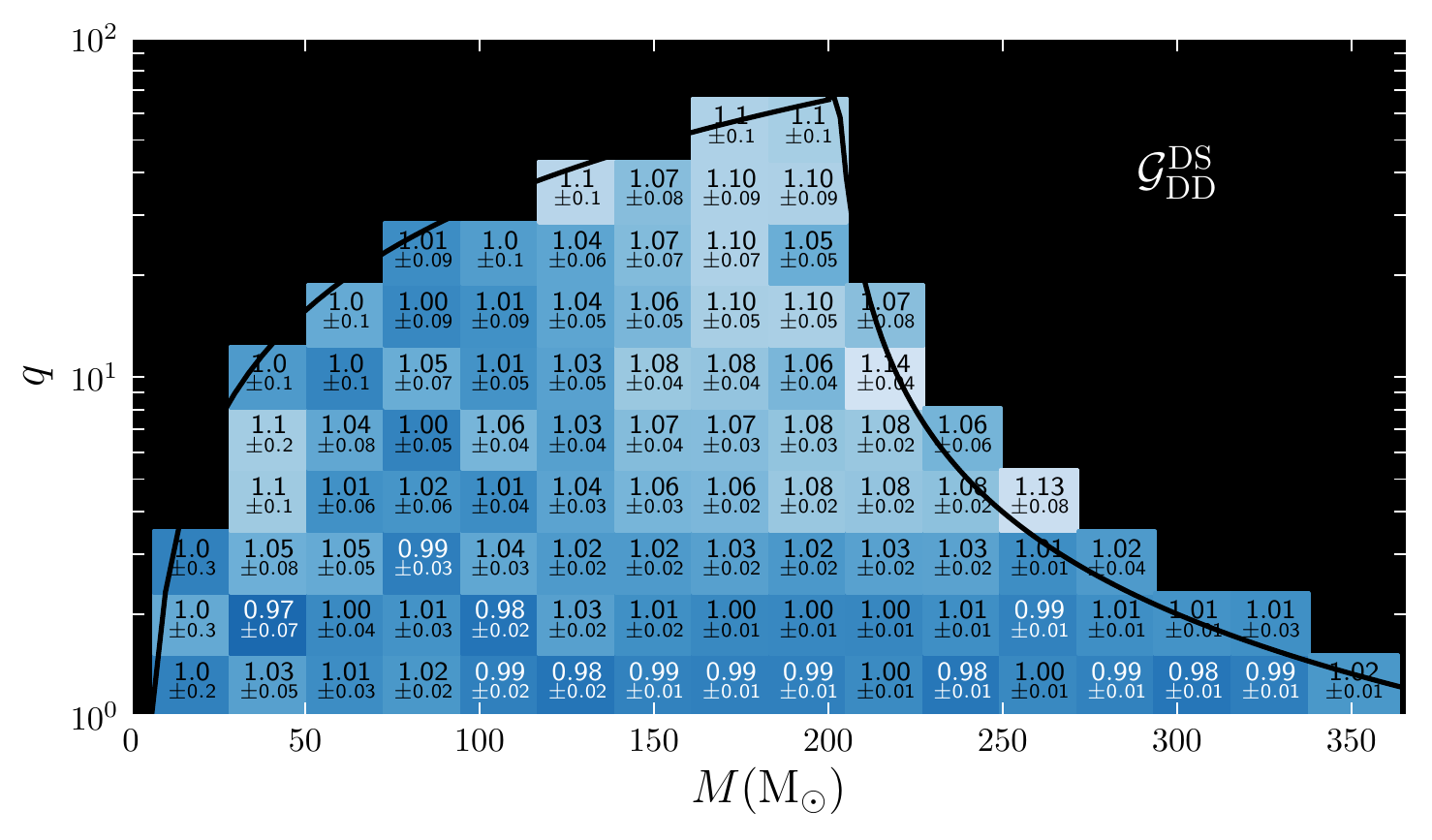}

\caption{Sensitive volume of injections with sub-dominant modes relative to
sensitive volume of injections without sub-dominant modes when both are
recovered by the dominant-mode template bank ($\gain{\dm\sd}{\dm\dm}$; see Sec.
\ref{ssec:statistics} for definition) in $M$, $q$ plane. The solid black line
indicates the injection region. Only tiles with $1000$ or more injections in
them are shown.} 

\label{fig:gainHM_DM}
\end{figure*}

Figure \ref{fig:gainHM_DM} shows $\gain{\dm\sd}{\dm\dm}$ in each mass bin. We
see that the gain $\approx 1$ everywhere. Indeed, if we assume a uniform
astrophysical rate in component masses we find $\netgain{\dm\sd}{\dm\dm} =
1.005 \pm 0.004$; a uniform rate in $M$ and $q$ yields
$\netgain{\dm\sd}{\dm\dm} = 1.019 \pm 0.004$. This means that a dominant-mode
template bank yields predicted detection rates even though both the bank and
the predictions neglect sub-dominant modes.

Comparing Fig. \ref{fig:gainHM_DM} to Fig. \ref{fig:HMeff} we see that
$\gain{\dm\sd}{\dm\dm} \approx 1$ despite the drop in effectualness to
injections with sub-dominant modes. Notably, the relative gain is slightly
\emph{larger than} $1$ in the areas where the effectualness is \emph{smallest}.
This means that the drop in effectualness seen in Fig. \ref{fig:HMeff} is
largely due to an increase in the amount of power in injections when
sub-dominant modes are added rather than a decrease in SNR between the
templates and injections.

\begin{figure}
\includegraphics[width=\columnwidth]{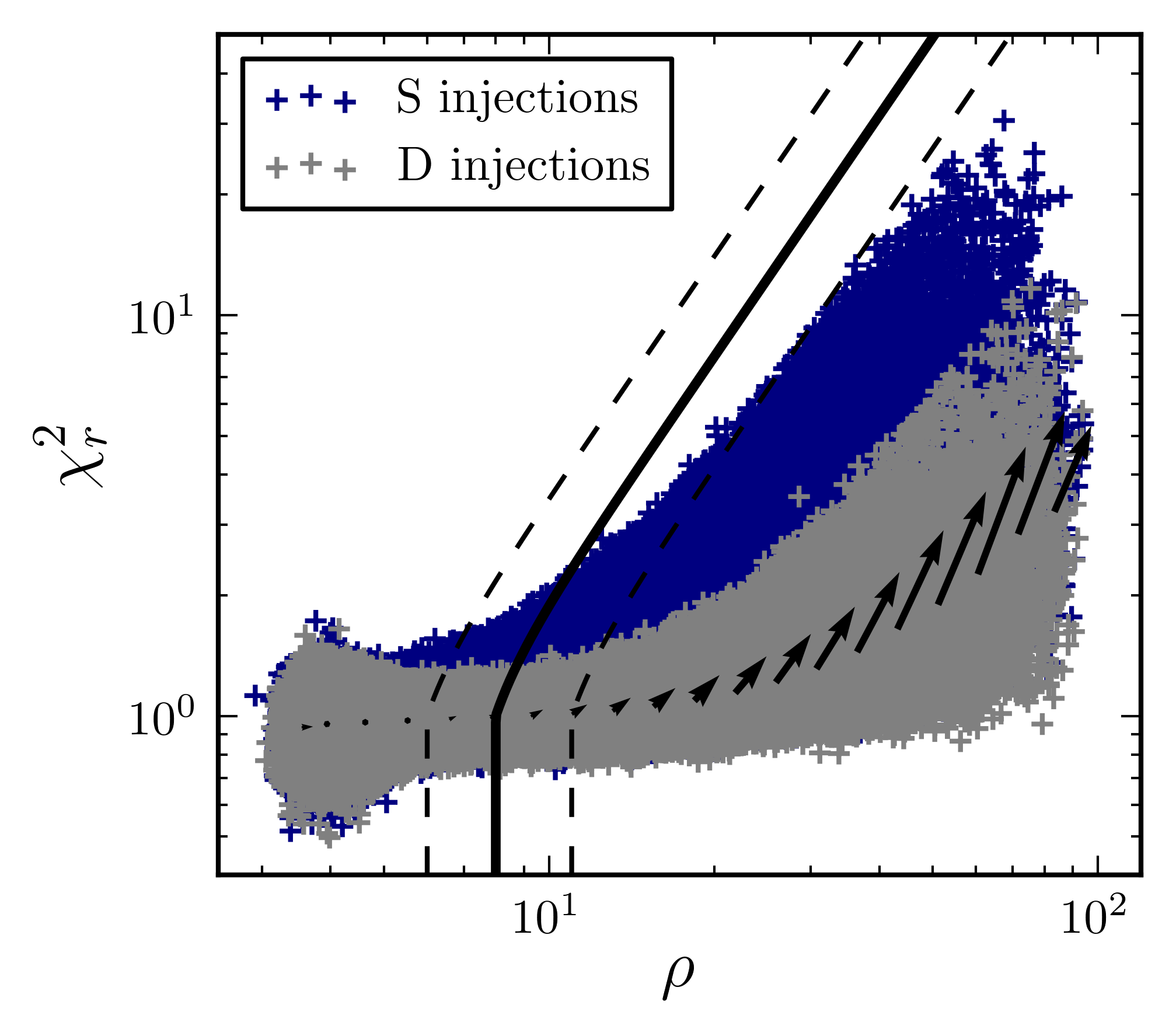}

\caption{Reduced $\chi^2$ ($\chi_r^2$) versus SNR of all injections with $M <
360\Msun$. Gray crosses are injections with the dominant-mode only, dark-blue
crosses are injections with sub-dominant modes. The dashed and sold black lines
show lines of constant $\rhonew$; shown are $\rhonew = 6,~8,$ and $11$ (dashed,
solid, and dashed, respectively). The black arrows indicate where, on average,
injections move in this plane when sub-dominant modes are added.}

\label{fig:snrChi}
\end{figure}

A decrease in effectualness will always result in an increase in $\chi^2$,
regardless of whether the mismatch ($\equiv 1 - \eff{}{}$) results in an
increase or decrease in SNR.  Whether re-weighted SNR increases or decreases
therefore depends on whether the gain in $\rho$ offsets the gain in $\chi^2$.
Figure \ref{fig:snrChi} shows reduced $\chi^2$ versus SNR for all injections
with $M<360\,\Msun$. The black arrows indicate how, on average, the injections
move in this plane when higher modes are added. The solid black line shows
where $\rhonew = 8$, which is the threshold we use when calculating efficiency.
All injections to the left of this line are missed, all injections to the right
are found. We see that at this threshold the increase in $\chi^2$ is roughly
offset by the increase in SNR when sub-dominant modes are added, which explains
why $\gain{\dm\sd}{\dm\dm} \approx 1$ despite an increase in mismatch for
injections with sub-dominant modes.

\begin{figure}
\includegraphics[width=\columnwidth]{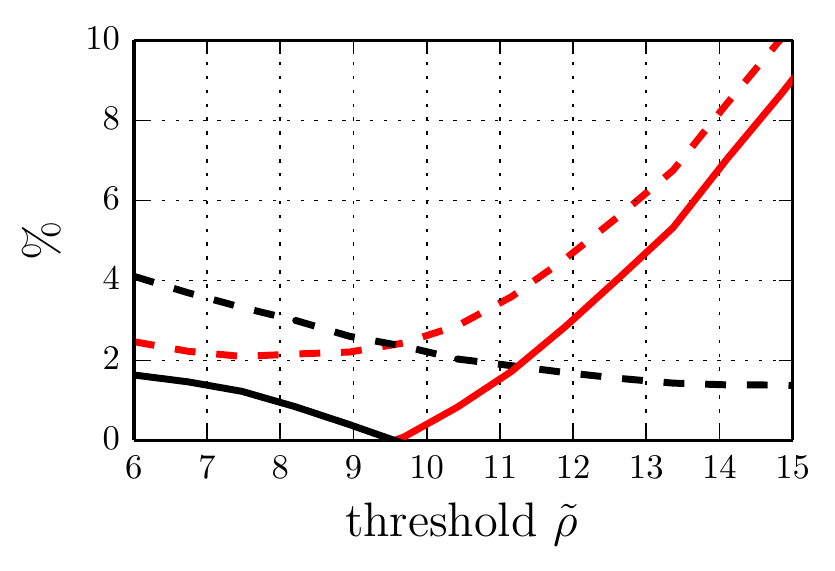} \\
\includegraphics[width=\columnwidth]{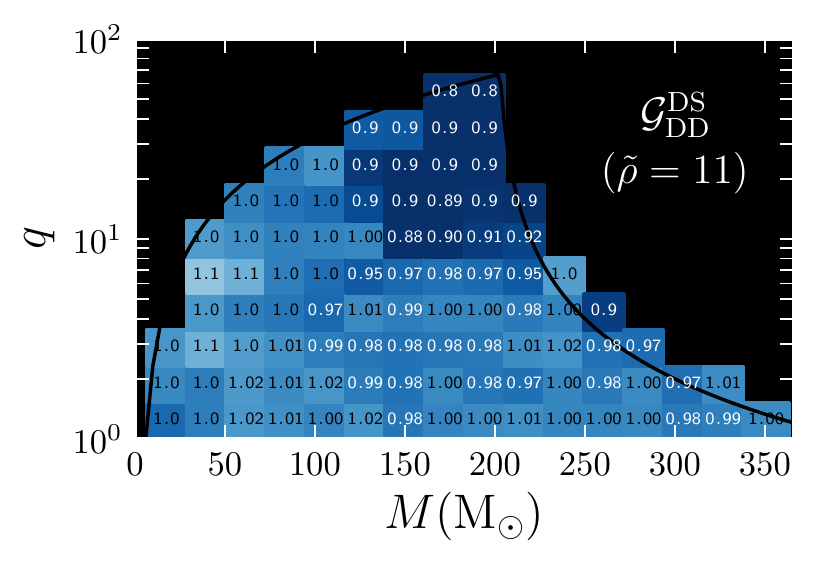}

\caption{\emph{Top}: Percentage of injections gained and lost when sub-dominant
modes are added as a function of threshold $\rhonew$. The dashed black line
shows the percentage of injections gained, the dashed red line shows the
percentage of injections lost. The solid line shows the net gain/loss; where it
is black there is a net gain; where it is red there is a net loss.
\emph{Bottom}: $\gain{\dm\sd}{\dm\dm}$ when the threshold re-weighted SNR for
both $\dm$ and $\sd$ injections is $11$ instead of $8$. The solid black line
indicates the injection region. Errors on gain are approximately twice the
errors reported in the corresponding tiles in Fig.  \ref{fig:gainHM_DM}.}

\label{fig:injFMratio}
\end{figure}

Figure \ref{fig:snrChi} also shows that as we go to higher $\rho$, the gain in
$\chi^2_r$ becomes larger relative to the gain in SNR when sub-dominant modes
are included (the black arrows progressively point to higher $\chi^2_r$ rather
than to higher SNR as we move to the right in the plot). If our threshold
$\rhonew$ were larger more injections would move below threshold (from an
increase in $\chi^2$) than above (from an increase in $\rho$). This is
confirmed by Fig.  \ref{fig:injFMratio}. The top plot shows the percentage of
injections that move above threshold (gained) and the percentage of injections
that move below threshold (lost) when sub-dominant modes are added, as a
function of threshold $\rhonew$. Around a threshold of $\rhonew = 8$ there is a
small net gain in the number of injections found when sub-dominant modes are
added. If the threshold were larger than $\sim 9.5$, however, there would be a
net loss.  This loss causes an overall drop in efficiency, which in turn causes
a drop in sensitivity. For example, the bottom plot in Fig.
\ref{fig:injFMratio} shows $\gain{\dm\sd}{\dm\dm}$ when the threshold $\rhonew
= 11$. We see that many tiles have gains $< 1$ now. Assuming a uniform
astrophysical rate in component-masses yields $\netgain{\dm\sd}{\dm\dm} = 0.994
\pm 0.004$; uniform in $M,q$ yields $\netgain{\dm\sd}{\dm\dm} = 0.988 \pm
0.006$.  Note, however, that the decrease in gain mostly occurs for tiles with
$M > 100\,\Msun$ and $q > 4$, which is where the drop in effectualness is the
largest. If we restrict our net gain calculation to stellar-mass BBHs only ($M
< 100\,\Msun$) we get $\netgain{\dm\sd}{\dm\dm} = 1.01 \pm 0.03$ for both rate
priors.

We conclude that if gravitational waves can be detected at a single-detector
$\rhonew \approx 9.5$ or less, then neglecting sub-dominant modes will not
affect the expected sensitivity of a non-spinning BBH search across all masses.
If the threshold for detection is larger, then neglecting sub-dominant modes
will cause a larger drop in sensitivity than what would otherwise be expected
for BBHs with $q \gtrsim 4$ and containing at least one IMBH ($m \gtrsim
100\,\Msun$).  Non-spinning stellar-mass BBHs, however, are not affected by
neglecting sub-dominant modes, even if the single-detector threshold for
detection is $\rhonew = 11$.

\section{Predicted efficiency of a template bank with sub-dominant modes}
\label{sec:effcyOfHMbank}

In the previous section we found that neglecting sub-dominant modes did not
adversely affect the sensitivity of a search for non-spinning BBHs. We found
this by comparing the sensitive volume of injections with sub-dominant modes to
the sensitive volume of injections without sub-dominant modes when both are
recovered by dominant-mode templates.  However, the decrease in effectualness
to injections with sub-dominant modes means that there is some power not being
recovered by the dominant-mode templates. If the templates had sub-dominant
modes they would be able to recover that power, which could increase the
sensitivity of the search. We therefore wish to know how the sensitivity of a
bank of templates with sub-dominant modes compares to a bank of templates
without sub-dominant modes, $\gain{\sd\sd}{\dm\sd}$.

\subsection{Search sensitivity when sub-dominant modes are included in all templates}

Currently, no search exists that uses a bank of templates with sub-dominant
modes. We do not try to create one here. Instead, we simulate how such a
search would perform as compared to the dominant-mode template bank used here.
In our simulation we will continue to use $\rhonew$ maximized over the template
bank as our detection statistic. We do not try to find the optimal search
statistic when sub-dominant modes are used.  As discussed in
Sec.~\ref{sssec:effectualness}, finding the optimal statistic over a bank of
templates is difficult, if not impossible, due to our lack of knowledge of the
mass distribution of BBHs. Our goal here is to answer the question: using
current search techniques, will sensitivity improve if we include sub-dominant
modes in templates?

To calculate $\gain{\sd\sd}{\dm\sd}$ we need to know the sensitive volume of a
template bank with sub-dominant modes to injections with sub-dominant modes,
$\vol{\sd}{\sd}$. Since $\vol{\sd}{\sd}$ depends on the number of injections
found by the template bank above some threshold, finding $\vol{\sd}{\sd}$
requires two pieces of information: the average re-weighted SNR of injections
filtered with a sub-dominant mode template bank $\avg{\rhonew_{\sd\sd}}$, and
the threshold for making a detection in such a search.

In order to estimate $\avg{\rhonew_{\sd\sd}}$ of each injection we first calculate
the maximum SNR that can be recovered from the injection by finding the overlap
of the injection with itself [see Eq. \eqref{eqn:maxSNR}]. A real sub-dominant
mode search will use discrete templates, as is done with the dominant-mode
bank. The discreteness of the template bank will cause some loss of SNR. Even
though both the injections and templates have sub-dominant modes in our
simulation, the mismatch due to the discreteness of the bank will also cause
some small increase in $\chi_r^2$, as was the case for the dominant-mode
injections recovered by the dominant-mode template bank (gray crosses in Fig.
\ref{fig:snrChi}). This in turn causes a small decrease in re-weighted SNR
relative to SNR for some injections. To simulate both of these effects, we
estimate $\avg{\rhonew_{\sd\sd}}$ to be:
\begin{equation}
\label{eqn:simulated_rhonew}
\avg{\rhonew_{\sd\sd}} = \frac{\avg{\rhonew_{\dm\dm}}}{\avg{\rho_{\dm\dm}}} \eff{\dm}{\dm}\max\{\rho_{\sd\sd}\}.
\end{equation}
Here $\eff{\dm}{\dm}$ simulates the loss in SNR due to the discreteness of the
bank --- i.e., we have assumed $\avg{\rho_{\sd\sd}} \approx
\eff{\dm}{\dm}\max\{\rho_{\sd\sd}\}$ --- and
$\avg{\rhonew_{\dm\dm}}/\avg{\rho_{\dm\dm}}$ simulates the effect of $\chi^2_r$
re-weighting. Thus the discreteness of our simulated template bank with
sub-dominant modes is equivalent to the discreteness of our (real)
dominant-mode bank.

By assuming $\avg{\rho_{\sd\sd}} \approx \eff{\dm}{\dm} \max\{\rho_{\sd\sd}\} =
\eff{\dm}{\dm}\sqrt{\ip{h_\sd}{h_\sd}}$ in Eq. \eqref{eqn:simulated_rhonew} we
have neglected any contribution to the SNR from noise. Due to the maximization
over the phase, the average SNR of a simulated signal $h$ when filtered with
itself in Gaussian noise will be slightly larger than $\sqrt{\ip{h}{h}}$.
Rather than adding a factor to our estimate of $\avg{\rhonew_{\sd\sd}}$ to account
for this contribution, when finding $\gain{\sd\sd}{\dm\sd}$ we use a
``noiseless" estimate of $\avg{\rhonew_{\dm\sd}}$, given by:
\begin{equation}
\avg{\rhonew_{\dm\sd}}^{\,\mathrm{(noiseless)}} = \frac{\avg{\rhonew_{\dm\sd}}}{\avg{\rho_{\dm\sd}}} \eff{\dm}{\sd} \max\{\rho_{\sd\sd}\}. 
\end{equation}
Here, $\eff{\dm}{\sd} \max\{\rho_{\sd\sd}\}$ gives $\left<\rho_{\dm\sd}\right>$
with the noise contribution removed and
$\avg{\rhonew_{\dm\sd}}/\avg{\rho_{\dm\sd}}$ accounts for $\chi^2_r$
re-weighting.

When calculating the sensitivity of the dominant-mode template bank we used a
threshold of $\rhonew = 8$. In a real search triggers are produced by
gravitational waves and by background noise. High statistical significance is
therefore required in order for a trigger to be considered a gravitational-wave
candidate. The standard measurement of significance is false-alarm probability
$\mathcal{F}(\rho)$. When evaluating the sensitivity of the dominant-mode
template bank we choose $\rhonew = 8$ because it corresponds to a false-alarm
probability small enough that we could confidently claim detection in real
detector data. This means that to compare the sensitivity of the simulated
sub-dominant mode search to the dominant-mode search we must chose a threshold
$\rhonew$ that results in the same false-alarm probability that $\rhonew = 8$
did in the dominant-mode search.

As discussed in Sec. \ref{ssec:DMsnr} sub-dominant modes break the degeneracy
between $\theta,~\phi$ and $\kappa$. This means that to fully recover all of
the power in the sub-dominant modes we have to maximize over three extrinsic
parameters instead of one. These additional maximizations will cause the
single-detector SNR distribution in noise to change. SNR from dominant-mode
templates is $\chi$ distributed with 2 degrees of freedom in stationary
Gaussian noise. If a template has sub-dominant modes, the SNR will have a
larger number of degrees of freedom. The increase in the degrees of freedom
means that $\mathcal{F}(\rho)$ at a given SNR will increase. In order to keep
the same false-alarm probability for a bank with sub-dominant modes as we had
for the dominant-mode bank, the threshold $\rhonew$ must therefore increase.
The relative gain in sensitivity of a template bank with sub-dominant modes
depends on whether the increase in SNR due to the sub-dominant modes is enough
to offset the increase in SNR threshold.

In Appendix \ref{appdx:bankFAP} we estimate the false-alarm probability of a
dominant-mode template bank with $N$ templates $\mathcal{F}(\rho |\dm, N)$ at
$\rho = 8$ [see Eq. \eqref{eqn:bankFAP_chi2}]. For the dominant-mode bank used
here $N = 19\,800$, which gives $\mathcal{F}(\rho = 8 | \dm, N) \approx 2.5
\times 10^{-10}$. We also calculate false-alarm probability as a function of
SNR of the simulated sub-dominant mode bank $\mathcal{F}(\rho | \sd)$ [see
Eqs. \eqref{eqn:bankFAP_general} and \eqref{eqn:intHMtemplatePDF}]. Assuming
$\rhonew \approx \rho$ for injections, we estimate that the threshold $\rhonew$
would have to increase to $8.31$ for the sub-dominant mode bank in order for
$\mathcal{F}(\rho | \sd, N)$ to also be equal to $2.5 \times 10^{-10}$.

\begin{figure*}
\begin{centering}
\includegraphics[width=2\columnwidth]{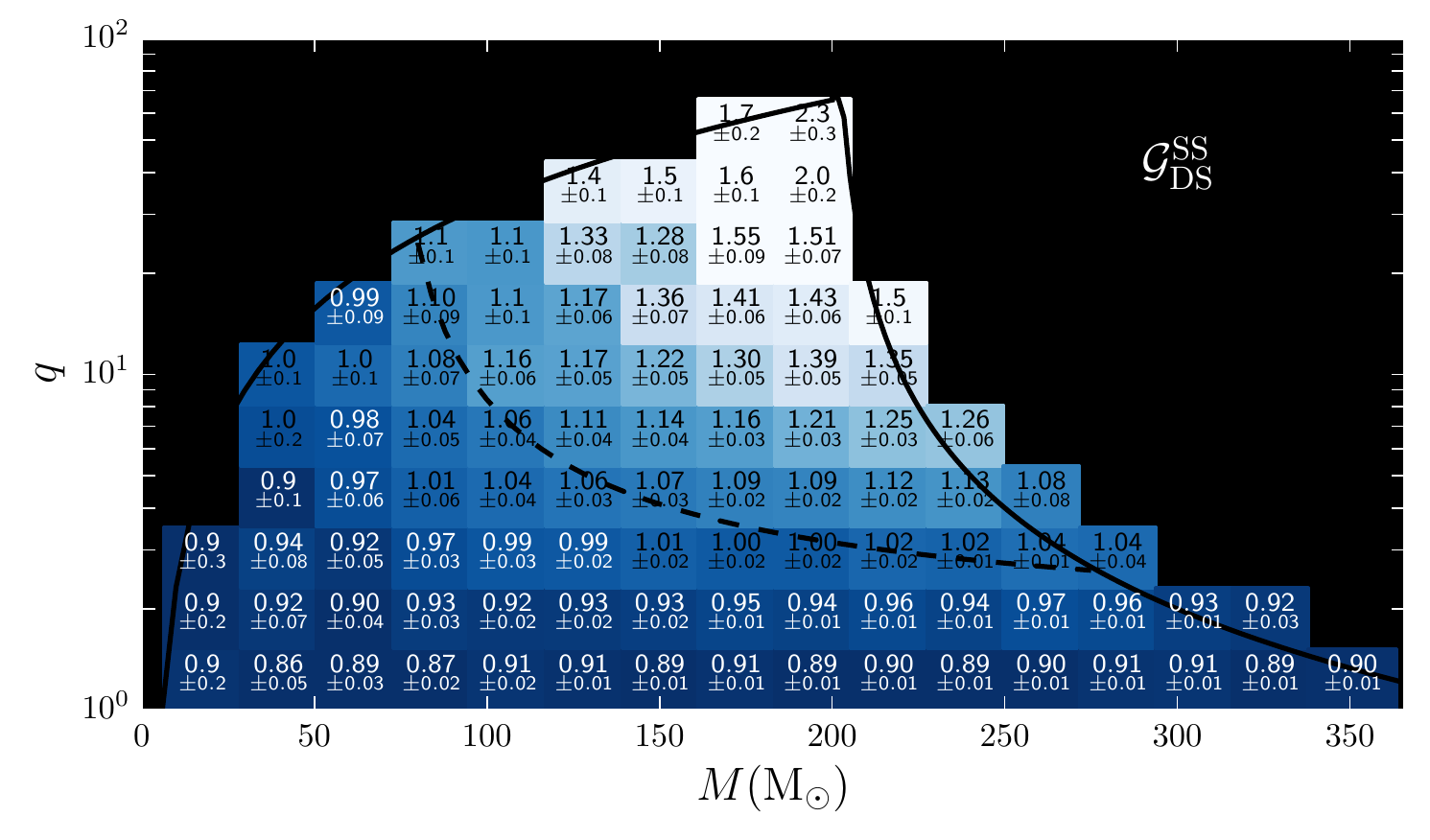} \\

\caption{Relative gain of sub-dominant bank compared to dominant-mode bank if
the sub-dominant modes are used everywhere, $\gain{\sd\sd}{\dm\sd}$. The solid
black line indicates the injection region. The dashed black line shows where we
choose to place a boundary between dominant-mode and sub-dominant mode
templates for a split-bank search (see Fig.  \ref{fig:gain-HMbank_cut} for
gain).}

\label{fig:gain-HMbank}
\end{centering}
\end{figure*}

Using the estimated $\rhonew_{\sd\sd}$ given in Eq.
\eqref{eqn:simulated_rhonew} and a threshold $\rhonew = 8.31$ we calculate the
sensitive volume of the simulated sub-dominant mode template bank $V_{\sd\sd}$
in the same $M,q$ tiles we used in Fig. \ref{fig:gainHM_DM}. We compare this to
$V_{\dm\sd}$ (which used a threshold $\rhonew = 8$) to get
$\gain{\sd\sd}{\dm\sd}$; this is plotted in Fig. \ref{fig:gain-HMbank}. We find
that the sub-dominant mode template bank has \emph{worse} sensitivity then the
dominant-mode bank across nearly the entire mass-space, dropping to as low as
$\sim 90\%$ for mass ratios between 1 and 1.5.

The reason for the poor sensitivity of the sub-dominant mode bank to equal-mass
systems can be understood by considering Fig. \ref{fig:HMeff}, in which we see
that when $q \rightarrow 1$, $\eff{\dm}{\sd} \rightarrow 1$. This means that
dominant-mode templates are recovering nearly all of the available power in
equal-mass signals. Put another way, signals from equal-mass systems gain
almost no SNR by adding sub-dominant modes to the templates. The SNR
distribution in noise of sub-dominant mode templates that are close to the
equal-mass line are not much different than their dominant-mode counterparts.
However, false-alarm probability is a \emph{global} property of the template
bank: even though the SNR distribution does not change for equal-mass
templates, it does for more asymmetric-mass templates. The presence of these
templates in the bank affects the false-alarm probability of the entire search.
The result is that equal-mass signals suffer an increase in threshold but gain
no SNR from the sub-dominant template bank.

The sub-dominant mode bank does have better sensitivity at higher mass ratios
and total masses. In the highest mass-ratio tile in Fig.  \ref{fig:HMeff} the
sensitivity is twice that of a dominant-mode bank. A sub-dominant mode bank may
therefore still increase the probability of making a detection, if the total
gain in sensitive volume to higher mass-ratio systems is enough to offset the
drop in sensitivity to equal-mass systems. However, assuming either a uniform
astrophysical rate in $m_1,m_2$  or a uniform rate in $M,q$ yield net gains $<
1$: the former gives $\netgain{\sd\sd}{\dm\sd} = 0.932 \pm 0.003$ while the
latter gives $\netgain{\sd\sd}{\dm\sd} = 0.987 \pm 0.003$. Restricting to
stellar-mass BBHs also results in net gains $< 1$, yielding $0.92 \pm 0.01$ and
$0.96 \pm 0.01$ for the two priors, respectively.

Using templates that more accurately model signals should improve sensitivity.
The reason the sub-dominant mode templates do not is
due to the astrophysical prior that is inherent in the search. Simply selecting
the template with the largest SNR when maximizing over the bank (as we have
done for both the dominant-mode and sub-dominant mode bank, and as is done in
current searches) assumes that every template is equally likely to detect a
signal. The detectors are not equally sensitive to all signals, however, nor is
the density of templates uniform. This makes the search most sensitive to a
particular astrophysical rate distribution. Adding sub-dominant modes changes
the distribution to which the search is most sensitive, thereby implicitly
changing the search prior.

\begin{figure}
\includegraphics[width=\columnwidth]{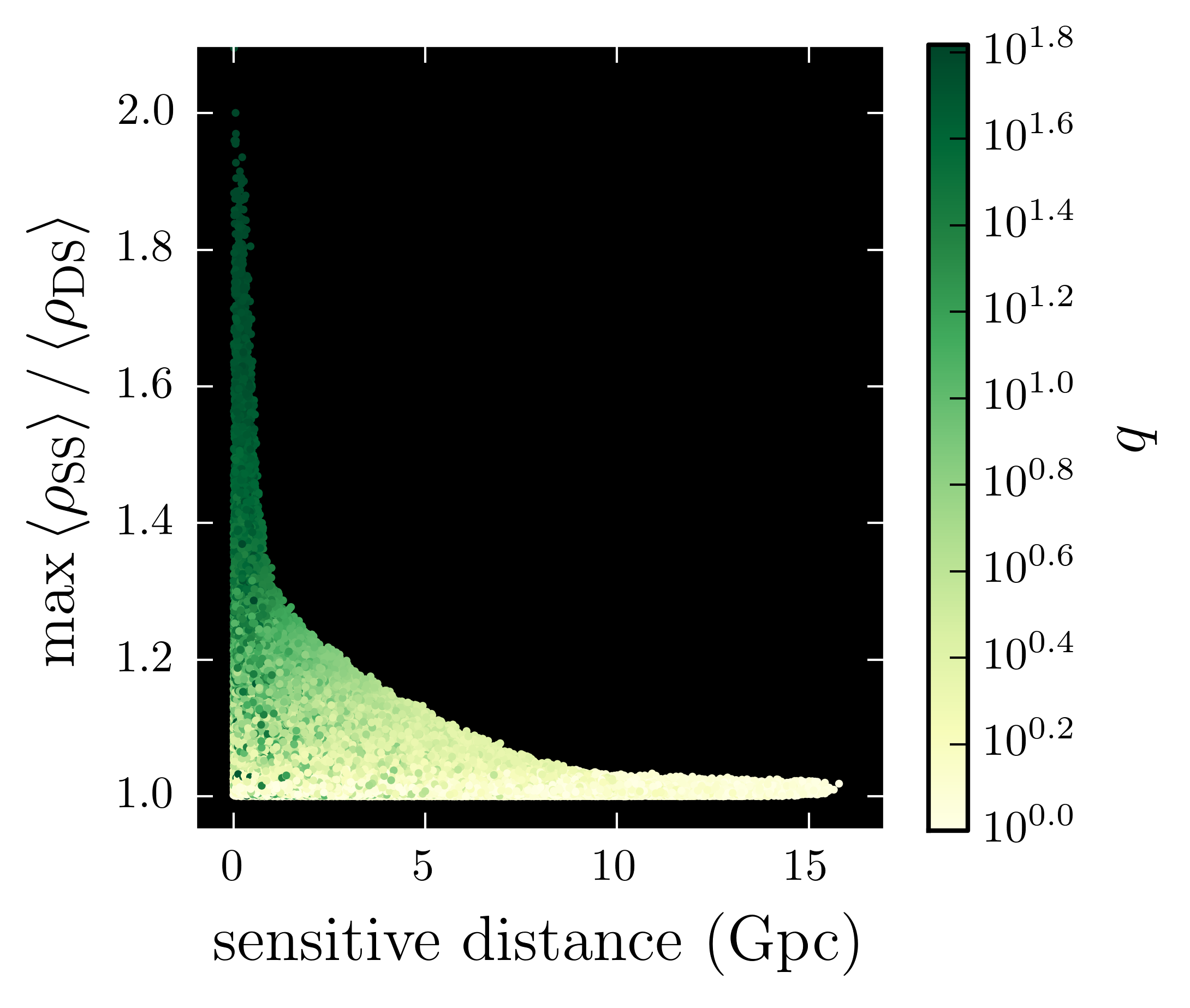} \\

\caption{The maximum gain in SNR that is possible by adding sub-dominant modes
to templates ($\max \avg{\rho_{\sd\sd}}/\avg{\rho_{\dm\sd}}$) versus the
distance at which a signal can be detected by the dominant-mode template bank
at SNR 8 (sensitive distance). Each point represents a simulated signal; the
points are colored by their mass-ratio ($q$).}

\label{fig:distGain}
\end{figure}

As noted by Refs. \cite{Pekowsky:2012sr} and \cite{Brown:2013hm}, the largest
SNR increase occurs for signals from asymmetric-mass binaries and from systems
inclined to the line of sight. Yet the magnitude of the detector's sensitivity
is lowest to these systems. Including sub-dominant modes while weighting the
SNR of each template equally causes us to gain sensitivity to signals for which
we are least sensitive at the cost of losing sensitivity to signals for which
we are most sensitive. We can see this negative correlation in Fig.
\ref{fig:distGain}, which shows the fractional gain in SNR when sub-dominant
modes are added (ignoring the effects of $\chi_r^2$ re-weighting) versus the
distance at which a signal can be detected by the dominant-mode bank at SNR 8
(``sensitive distance"), colored by mass-ratio. The sensitive distance of the
signals with the largest gain (and mass-ratio) is an order of magnitude smaller
than the sensitive distance of the signals with the smallest gain. In general,
a plot such as this can be used to determine whether changing the parameters of
a search are worthwhile.  If there is a negative correlation between SNR gain
and sensitive distance --- i.e., only signals for which the detector is least
sensitive gain SNR by changing the search parameters  --- that change is
unlikely to improve the overall detection rate unless the astrophysical
distribution is strongly weighted to those signals.

\subsection{Splitting the template bank to improve sensitivity}

Including sub-dominant modes in templates would improve the sensitivity of the
search if we applied a weight to each template that better reflected the
probability the template will make a detection. Determining the best weight to
use requires knowledge of the astrophysical rate of signals. We expect that
signals will be distributed uniformly in the inclination angle $\theta$.
However, the proper weight to apply to signals with this assumption is already
inherent in the SNR. The projection of the waveform into the radiation frame,
and then again into the detector's frame (done by the $\mtrx{P}$ and $\mtrx{K}$
matrices, respectively, in Appendix \ref{appdx:maxHMsnr-analytic}) \emph{is}
the weight one should apply to the SNR assuming signals are distributed
uniformly in space. Indeed, that the SNR from an inclined signal is smaller
than the SNR from a face-on signal at the same distance is a result of the
antenna pattern indicating that these signals are less likely to be detected.

Any weight we apply to the SNR must therefore be based on the distribution
of intrinsic paramters. Given the large uncertainty in the mass distribution
of BBHs this is difficult to do. However, if we can find a weight that results
in $\gain{}{} \geq 1$ relative to the dominant-mode search in all mass tiles,
we know that this new search will be at least as sensitive to real signals as
the dominant-mode search, regardless of the astrophysical mass distribution.

The simplest way to arrive at such a weight is to split the template bank
in two parts. In one part we use templates with sub-dominant modes; in the
other, we use dominant-mode templates. In this scenario false-alarm
probabilities are calculated separately in each region, then combined. As we
will see below, the process of combining the results across the split
effectively down-weights higher mass-ratio templates such that they do not hurt
the sensitivity to equal-mass systems, yet still improves sensitivity to
the high-mass ratio signals.

In order to not lose any sensitivity we need to place the split between
dominant-mode and sub-dominant mode templates at points in the mass space where
$\gain{\sd\sd}{\dm\sd} > 1$. Otherwise, signals with masses that are in the
sub-dominant part of the bank that are near the split will suffer a loss of
sensitivity. We empirically chose to split the bank such that sub-dominant mode
templates are used when:\footnote{We put the cut in $m_1, m_2$ rather then $M,
q$ because we found that the $\gain{\sd\sd}{\dm\sd}$ was roughly linear in
$m_1, m_2$.}
\begin{equation}
\label{eqn:split_bank}
m_2 < 0.6 m_1 - 43.
\end{equation}
This cut is shown in Fig. \ref{fig:gain-HMbank} (black dashed line). We note that
this is an empirical estimate for this template bank and Gaussian noise. If a
real search were performed, this cut would have to be tuned based on the noise
and bank characteristics. We assume that such a cut would be in approximately
the same location in parameter space as the cut we choose here, however. 

Using the split in Eq. \eqref{eqn:split_bank} we find that $3750$ templates are
in the sub-dominant region while $16\,050$ templates remain in the
dominant-mode region. In Appendix \ref{appdx:bankFAP} we find that
$\mathcal{F}(\rho)$ is roughly proportional to the number of templates in the
search [see Eq. \eqref{eqn:bankFAP_general}]. This means that the false-alarm
probabilities we get for triggers in the dominant-mode part of the template
bank will have decreased by a factor of $16\,050/19\,800 \approx 0.8$ compared
to what we get when we search the entire mass space. Likewise, the false-alarm
probability of triggers in the sub-dominant mode part of the bank will decrease
by a factor of $\sim 0.2$.  These drops in false-alarm probabilities are
artificial: they are due solely to our choice of cut. Indeed, we could split
the bank an arbitrary number of times, thereby arbitrarily decreasing
false-alarm probabilities.

To account for this artificial decrease we must combine results by multiplying
the signals' false alarm probabilities by a \emph{trials factor}
\cite{\sSixLowMass}. For the dominant-mode part of the bank we multiply by
$19\,800/16\,050 \approx 1.2$; this is equivalent to using dominant mode
templates throughout the entire mass space, and so the threshold in this part
of the bank remains $\rhonew = 8$. In the sub-dominant mode part of the bank we
multiply by $19\,800/3750 \approx 5.3$. To keep $\mathcal{F}(\rho)$ fixed at
$2.5 \times 10^{-10}$, the threshold in this region increases to $8.44$. This
is equivalent to down-weighting $\rhonew$ of templates in this part of the
bank, as desired.

\begin{figure*}
\begin{centering}
\includegraphics[width=2\columnwidth]{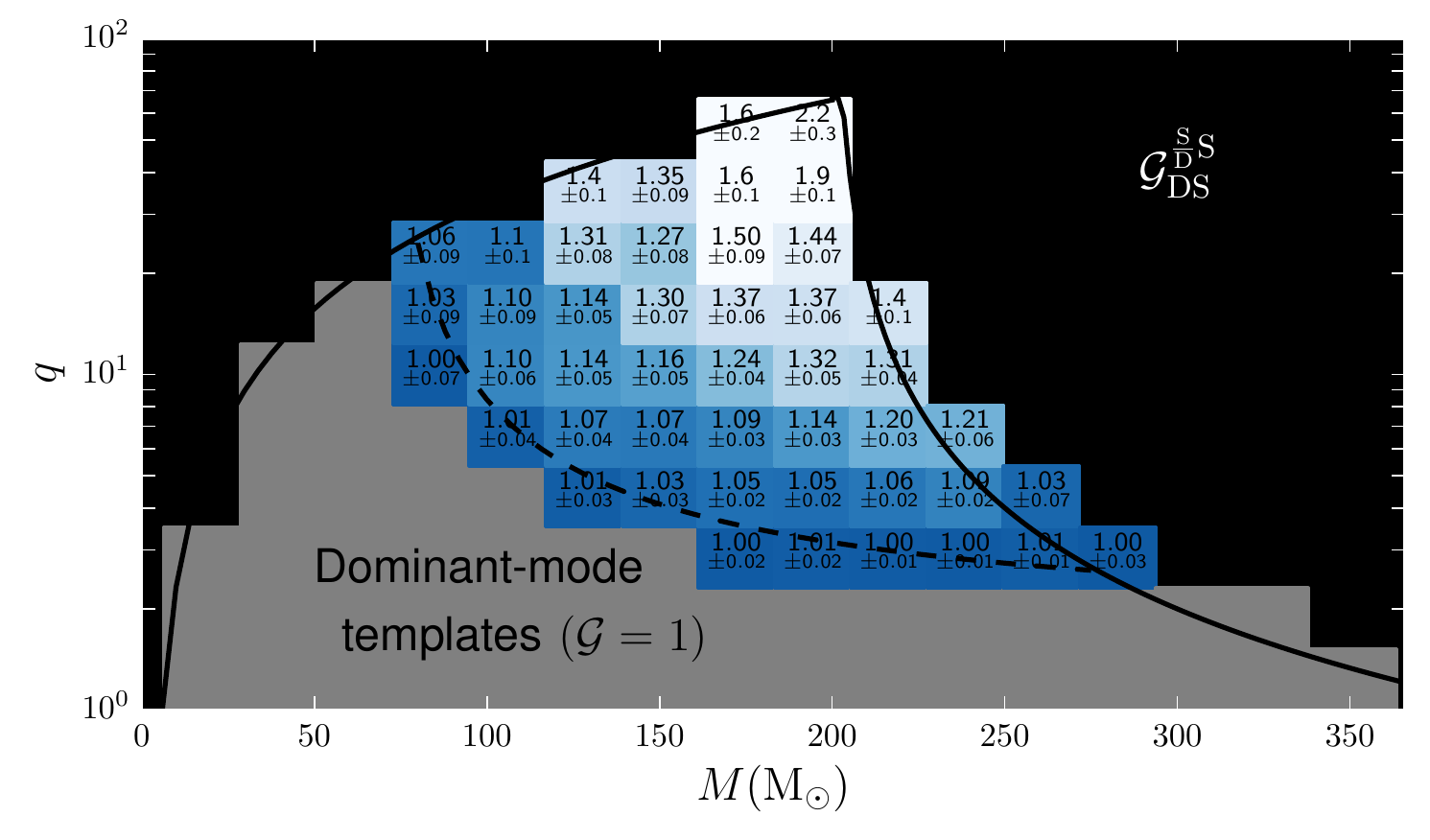}

\caption{Relative gain of sub-dominant bank compared to dominant-mode bank if
the sub-dominant modes are only used in part of the bank,
$\gain{\frac{\sd}{\dm}\sd}{\dm\sd}$. The sold black line indicates the
injection region; the dashed-black line indicates the boundary used for
switching between dominant-mode and sub-dominant mode templates.}

\label{fig:gain-HMbank_cut}
\end{centering}
\end{figure*}

Figure \ref{fig:gain-HMbank_cut} shows the relative gain of the split bank
(indicated by $\frac{\sd}{\dm}$) compared to using dominant mode templates
everywhere ($\gain{\frac{\sd}{\dm}\sd}{\dm\sd}$). We find that we can gain
sensitivity by using templates with sub-dominant modes if we split the bank
using the criteria in Eq.  \eqref{eqn:split_bank}. Since dominant-mode
templates are used below this cut, we lose no sensitivity relative to what we
had when we used dominant-mode templates everywhere.  In terms of possible
astrophysical systems, note that only IMRIs ($m_1 \gtrsim 100\,\Msun, ~m_2
\lesssim 80\,\Msun$) fall in the region where sub-dominant modes are used, and
that the dominant-mode part of the bank covers all possible stellar-mass BBHs
($m_1,m_2 \lesssim 80\,\Msun$).

\begin{figure*}
\begin{centering}
\includegraphics[width=2\columnwidth]{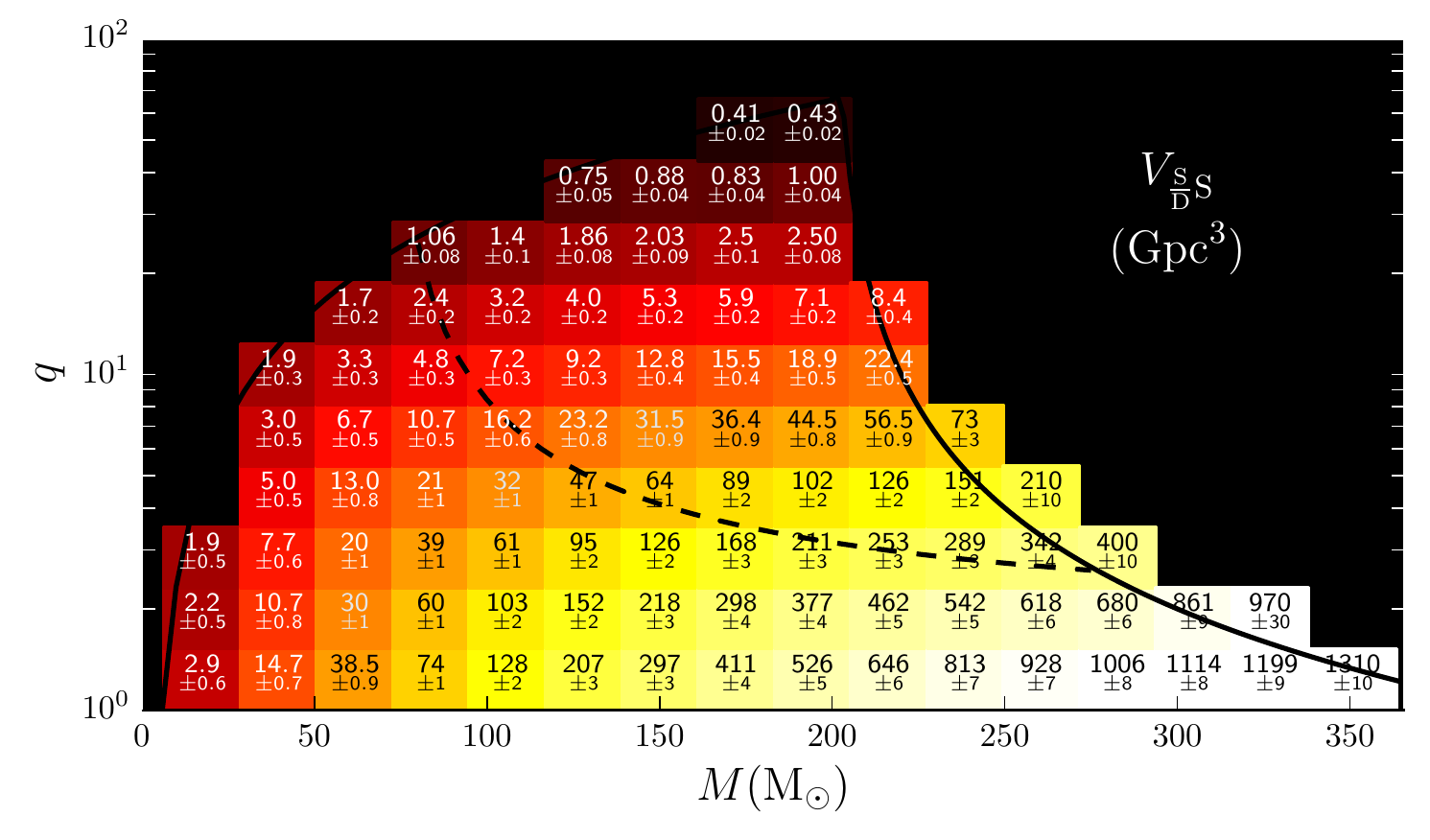}

\caption{Sensitive volume in $\mathrm{Gpc}^3$ of the split bank. The
solid-black line shows the injection region, the dashed-black line the cut
between sub-dominant and dominant-mode templates. Dividing the sensitive
volumes in this figure by the net gains given in Fig. \ref{fig:gain-HMbank_cut}
yields sensitive volumes if dominant-mode templates are used everywhere.}

\label{fig:sensitiveVol}
\end{centering}
\end{figure*}

Splitting the bank and only using sub-dominant modes in one part results in a
search that is as good or better than the dominant-mode search in all mass
tiles, as we desired. The net gain, however, is not much larger than simply
using dominant-mode templates everywhere. Assuming a uniform astrophysical rate
in component masses gives $\netgain{\frac{\sd}{\dm}\sd}{\dm\sd} = 1.002 \pm
0.004$; assuming a uniform rate in $M,q$ yields $1.024\pm 0.004$. The reason
for the nearly non-existent increase in net gains can be understood by
considering Fig. \ref{fig:sensitiveVol}, which shows the sensitive volume of
the split bank in Gpc$^3$. For a given total mass, the largest sensitivity
occurs at equal mass by a large margin. For example, between $\sim 190$ and
$200\,\Msun$ (the mass bin with the largest range in $q$) we see that the
sensitive volume drops by four orders of magnitude --- from
$526\,\mathrm{Gpc}^3$ to $0.43\,\mathrm{Gpc}^3$ --- as we go from $q = 1$ to $q
\approx 60$. Even though the sub-dominant modes have doubled our sensitivity at
the highest mass ratio tile, the magnitude of the sensitive volume is small
relative to the other parts of the bank. Using the split bank is therefore
unlikely to have a significant affect on the overall probability of making a
gravitational-wave detection in advanced LIGO, unless there is a relatively
large population of IMRIs with $q \gtrsim 4$ compared to IMBH-IMBH and
stellar-mass BBHs.

\section{Conclusions and Future Work}
\label{sec:conclusions}

We have investigated the effects of neglecting sub-dominant modes in templates
on the predicted BBH sensitivity of a single advanced LIGO detector. In doing
so we have considered the loss in re-weighted SNR, which more accurately
reflects what is done in real BBH searches than considering SNR alone. We found
that not including sub-dominant modes in templates does not make the
sensitivity to non-spinning BBH signals any worse than what has been predicted
as long as the gravitational waves can be detected at single-detector
re-weighted SNRs $\lesssim 9.5$. If the threshold for detection is larger,
neglecting sub-dominant modes will result in worse sensitivity than predicted,
but only for IMRI BBHs with total masses $\gtrsim 100\,\Msun$ and $q \gtrsim
4$. Stellar-mass BBHs ($M \lesssim 100\,\Msun$) and more equal-mass IMBH-IMBH
binaries ($m_{1,2} \gtrsim 100\,\Msun, ~q \lesssim 4$) are unaffected.

We have also simulated a bank of non-spinning templates that have sub-dominant
modes to see if any improvement in sensitivity can be gained. To do so we
analytically maximized the SNR over $\kappa$ when sub-dominant modes are
included, then numerically maximized over the remaining extrinsic parameters
[see Appendix \ref{appdx:maxHMsnr-analytic}]. We found that if sub-dominant
modes are used throughout the entire bank, and all templates are weighted
equally, the sensitivity would be \emph{worse} for stellar-mass BBHs and
IMBH-IMBH binaries with $q \lesssim 4$ due to the increase in threshold needed
to keep false-alarm probability fixed.  Such a search would only improve
sensitivity to IMRIs with $q \gtrsim 4$. Therefore, in order to include
sub-dominant modes in templates, the templates must be weighted by the
likelihood that they will detect a signal.

We found an effective weight by splitting the bank in two parts. Using
sub-dominant modes in templates that satisfy Eq. \eqref{eqn:split_bank} and
dominant-mode templates elsewhere, we can improve the sensitivity to IMRI BBHs
without sacrificing the sensitivity to stellar-mass BBHs and IMBH-IMBH
binaries. The split bank is only one of many possible weighting schemes.
Using sub-dominant modes throughout the entire bank with a more incremental
weight may yield better sensitivity. However, since sub-dominant modes improve
sensitivity to signals for which the detectors are least sensitive, a more
sophisticated search would probably only yield a small increase in the
detection rate.

In Fig. \ref{fig:sensitiveVol} we present the single-detector sensitive volumes
of the split bank. We find that the sensitive volumes to signals in the
sub-dominant part of the bank (tiles above the black-dashed line) are one to
four orders of magnitude smaller than equal mass-ratio systems with equivalent
total masses. Using the split-bank is therefore unlikely to substantially
increase the probability of making a gravitational-wave detection over a
dominant-mode bank unless there is a large population of high-mass ratio IMRIs
in the universe.

In order to use a split bank more investigation is required to establish how
exactly to use sub-dominant modes in a real search.  Open questions include how
to search over $\theta$ and $\phi$,\footnote{One possibility is to simply place
templates in $\theta$ and $\phi$ using the stochastic method described in Ref.
\cite{2012arXiv1210.6666A} In that case, the SNR of each template would be
found using Eq. \eqref{eqn:SNRmaxK}.} and how to apply a coincidence test
between multiple detectors. Given that using a split bank has negligible impact
on the overall probability of making a gravitational-wave detection in advanced
LIGO, simply using a dominant-mode bank everywhere may be more desirable.

We did not try to predict advanced LIGO BBH detection rates, as doing so would
require a choice of astrophysical rates. However, Fig. \ref{fig:sensitiveVol}
can be used to predict detection rates if a particular astrophysical rate is
assumed. The volumes given in Fig. \ref{fig:sensitiveVol} are for a split bank;
dividing the sensitive volumes by the net gains given in Fig.
\ref{fig:gain-HMbank_cut} yields the sensitive volumes if a dominant-mode bank
is used everywhere instead. For example, if a dominant-mode bank is used, the
sensitive volume of the largest-mass ratio tile is $\sim 0.2\,\mathrm{Gpc}^3$
instead of $0.43\,\mathrm{Gpc}^3$. The sensitive volumes we report were
calculated using a single detector. Since real searches use a network of
detectors, actual sensitive volumes may vary depending on the relative
sensitivities of each detector.

We emphasize that in this study we only considered non-spinning signals.
Sub-dominant modes are likely to play a more important role when one or both of
the component masses are spinning.  Currently, there are no spinning waveform
models available with merger and ringdown that include sub-dominant modes.
Once such waveforms become available, creating a sub-dominant mode search may
be more advantageous. Since our analytic maximization over $\kappa$ in Appendix
\ref{appdx:maxHMsnr-analytic} is still valid if the component masses are
spinning, the result therein [specifically Eq. \eqref{eqn:SNRmaxK}] can be used
in such a search.

A dominant-mode EOB model calibrated to numerical relativity that incorporates
spins aligned with the orbital angular momentum does currently exist
\cite{Taracchini:2012}, as well as spinning ``phenomological" models derived
from numerical relativity \cite{Ajith:2009bn,Santamaria:2010}.  Past BBH
searches have only used non-spinning templates, but there is much work
currently on-going to extend template banks into the spinning regime
\cite{2012arXiv1210.6666A, Brown:2012tb}. Doing so brings up many of the same
questions we addressed in this study: is the sensitivity of a search
significantly affected by neglecting spin?  Does the increase in number of
templates needed to include spin increase the false-alarm probability such that
any gain in SNR is nullified?  In a future work we will address these questions
using spinning EOB waveforms calibrated to numerical relativity and the methods
that we establish in this study.

\begin{acknowledgments} 

We are grateful to Kipp Cannon, Thomas Dent, Prayush Kumar, Cole Miller,
Alex Nitz, and John Whelan for valuable discussions. We would also like to
thank Duncan Brown for providing some of the software used in this analysis.
A.B., C.C. and Y.P.  acknowledge partial support from NSF Grants No.
PHY-0903631 and No.  PHY-1208881.  A.B. also acknowledges partial support from
the NASA Grant NNX12AN10G. All calculations were performed on the SUGAR
cluster, which is supported by NSF Grants No.  PHY-1040231, No.  PHY-1104371,
and No.  PHY-0600953, and by Syracuse University ITS.

\end{acknowledgments}

\appendix

\section{SNR maximization with sub-dominant modes}
\label{appdx:maxHMsnr-analytic}

All of the assumptions we made for dominant-mode templates apply to templates
with sub-dominant modes up to Eq. \eqref{eqn:simplifiedSNR}. As stated in Sec.
\ref{sec:background}, when only the dominant-mode is considered, the angles
$\kappa, ~\theta$, and $\phi$ can be combined into a single parameter.  The
maximization over this parameter is straight forward as the denominator of the
SNR does not depend on it (the parameter cancels in the inner product
$\sqrt{\ip{h}{h}}$) \cite{Sathyaprakash:1991mt}.  However, when sub-dominant
modes are considered, the maximization over the extrinsic parameters is more
complicated.  First, $\kappa, ~\theta$, and $\phi$ cannot be combined as the
degeneracy between them is broken. Second, both the numerator and the
denominator in the SNR depend on all three angles [see Eq.
\eqref{eqn:matrix_SNR}, below]. Here we derive an analytic expression for the
SNR maximized over $\kappa$ when all modes are included in the waveform. As the
result [Eq. \eqref{eqn:SNRmaxK}] has non-trivial dependence on $\theta$ and
$\phi$ we do the remaining maximizations numerically.

The $m$ summation in Eq. \eqref{eqn:hplus_hcross} is over $-l \leq m \leq l$.
The number of terms in the summation can be reduced by relating the positive
and negative $m$ modes. Letting:
\begin{align*}
{}_{-2}Y_{lm}(\theta, \phi) &= g_{lm}(\theta)e^{i m \phi}, \\
h_{lm}(t) &= A_{lm}(t)^{-i m \Psi_{lm}(t)},
\end{align*}
$h$ is:
\begin{align}
\label{eqn:pos_neg_m}
h(t) &= \sum_{l|m|} \Big(\Re\big\{g_{lm}(\theta) e^{i m \phi} h_{lm}(t) \nonumber \\
     &\qquad \qquad + g_{l-m}(\theta) e^{-i m \phi} h_{l-m}(t)\big\}\Big) \cos\kappa \nonumber \\
    &\quad + \Big(\Im\big\{g_{lm}(\theta) e^{i m \phi} h_{lm}(t) \nonumber \\
    &\qquad \qquad + g_{l-m}(\theta) e^{-i m \phi}h_{l-m}(t)\big\}\Big) \sin\kappa.
\end{align}
Here, $|m|$ indicates that the summation is over positive-$m$ modes only, and we
have set $\mathcal{D} = 1$ since it does not enter in the SNR.\footnote{As
written, Eq. \eqref{eqn:pos_neg_m} double counts terms for which $m=0$. For the
sake of brevity we have redefined the $A_{l0}$ in this equation to be $1/2$ of
their original values.} The $\pm m$ modes of the $h_{lm}$ are related by:
\begin{equation}
\label{eqn:pos_neg_hlm}
h_{l-m} = (-1)^l h^{*}_{lm}.
\end{equation}
Using this relationship Eq. \eqref{eqn:pos_neg_m} becomes:
\begin{align}
\label{eqn:h_pre_matrix}
h(t) &= \sum_{l|m|} \Big( G_{1lm}(\theta) \cos(m\phi) h_{1lm}(t) \nonumber \\
     &\qquad \qquad + G_{1lm}(\theta)\sin(m\phi) h_{2lm}(t) \Big) \cos \kappa \nonumber \\
     &\quad + \Big( G_{2lm}(\theta) \sin(m\phi) h_{1lm}(t) \nonumber \\
     &\qquad \qquad - G_{2lm}\cos(m\phi) h_{2lm}(t)\Big) \sin \kappa,
\end{align}
where:
\begin{align}
\label{eqn:G1lm}
G_{1lm}(\theta) &= g_{lm}(\theta) + (-1)^l g_{l-m}(\theta); \\
\label{eqn:G2lm}
G_{2lm}(\theta) &= g_{lm}(\theta) - (-1)^{l} g_{l-m}(\theta); \\
\label{eqn:h1lm}
h_{1lm}(t) &= A_{lm}(t) \cos[m\Psi(t)]; \\
\label{eqn:h2lm}
h_{2lm}(t) &= A_{lm}(t) \sin[m\Psi(t)].
\end{align}

Equation \eqref{eqn:h_pre_matrix} can be expressed more concisely by the matrix
equation:
\begin{align}
\label{eqn:h_matrix}
h(t) &= \sum_{l|m|} \mtrx{K} \mathcal{P}_{lm} \mathcal{H}_{lm} \\
     &= \mtrx{K} \mtrx{P} \mtrx{H},
\end{align}
where:
\begin{align}
\mathcal{P}_{lm} &=
    \begin{pmatrix}
    G_{1lm}\cos m\phi & G_{1lm}\sin m\phi \\
    G_{2lm}\sin m\phi & -G_{2lm}\cos m\phi
    \end{pmatrix}; \\
\mathcal{H}_{lm} &=
    \begin{pmatrix}
    h_{1lm} \\
    h_{2lm}
    \end{pmatrix}; \\
\mtrx{K} &= 
    \begin{bmatrix}
    \cos\kappa & \sin\kappa
    \end{bmatrix}; \\
\mtrx{P} &=
    \begin{bmatrix}
    \mathcal{P}_{22} & \mathcal{P}_{21} & \cdots
    \end{bmatrix}; \\
\mtrx{H} &=
    \begin{bmatrix}
    \mathcal{H}_{22} \\
    \mathcal{H}_{21} \\
    \vdots
    \end{bmatrix}.
\end{align}
Let us define the \emph{covariance matrix} $\mtrx{C}$ as:
\begin{equation}
\label{eqn:covar_matrix}
\mtrx{C} \equiv \mtrx{H} \otimes \mtrx{H},
\end{equation}
where the outer product $\mtrx{A}\otimes\mtrx{B}$ is:
\begin{equation}
[\mtrx{A} \otimes \mtrx{B}]_{ij} \equiv \ip{A_{i}}{B_{j}}.
\end{equation}
Note that $\mtrx{K}$ depends only on $\kappa$, $\mtrx{P}$ on $\theta$ and
$\phi$, and $\mtrx{C}$ on the intrinsic parameters $\intP$. Also note that
$\mtrx{C}$ is symmetric. In this notation $\ip{h}{h}$ is:
\begin{equation}
\label{eqn:matrix_hh}
\ip{h}{h} = \mtrx{K}\mtrx{P}\mtrx{C}\mtrx{P^T}\mtrx{K^T}.
\end{equation}
Defining $\mtrx{Q}$ as:
\begin{equation}
\label{eqn:QmatrixDef}
\mtrx{Q} = \mtrx{H}\otimes\begin{bmatrix}s\end{bmatrix} =
    \begin{bmatrix}
    \ip{h_{122}}{s} \\
    \ip{h_{222}}{s} \\
    \vdots
    \end{bmatrix},
\end{equation}
the SNR is:
\begin{equation}
\label{eqn:matrix_SNR}
\rho = \max_{\theta, \phi, \kappa} \frac{\mtrx{K}\mtrx{P}\mtrx{Q}}{\sqrt{\mtrx{K}\mtrx{P}\mtrx{C}\mtrx{P^T}\mtrx{K^T}}}.
\end{equation}

We can perform the maximization over $\kappa$ analytically. First consider the
case when $\theta = \pi/2$. From the definition of the ${}_{-2}Y_{lm}$:
\begin{equation*}
G_{2lm}(\pi/2) = 0, ~\forall (l,m),
\end{equation*}
which means:
\begin{equation*}
P_{2i}(\theta = \pi/2,\phi) = 0, ~ \forall i.
\end{equation*}
In this case $\mtrx{KP}$ has no $\sin\kappa$ terms, causing the $\kappa$
dependence to cancel in Eq. \eqref{eqn:matrix_SNR}. Thus\footnote{Here repeated
indices indicate sum over.}:
\begin{equation*}
\rho = \max_{\phi, \theta = \pi/2} \frac{P_{1i} Q_i}{\sqrt{P_{1j} C_{jk} P_{k1}}}.
\end{equation*}

To maximize over $\kappa$ when $\theta \neq \pi/2$, let $\mtrx{g} =
\mtrx{P}\mtrx{C}\mtrx{P^T}$ so that the denominator of the SNR is
$\sqrt{\mtrx{K g K^T}}$. Note that $\mtrx{g}$ is a $2\times2$ symmetric
matrix. We can therefore think of it as a metric that defines an inner product
space between 2-dimensional vectors. Since $\mtrx{g}$ depends on $\mtrx{P}$ and
$\mtrx{C}$ the curvature of this space is determined by $\theta,~\phi$ and
$\intP$. From this point of view the denominator of the SNR is simply the
magnitude of $\mtrx{K}$ in this space:\footnote{We use $\arbip{\cdot}{\cdot}$
to differentiate the inner product defined by $\mtrx{g}$ from the inner product
defined in Eq. \eqref{eqn:defIP}.}
\begin{equation*}
\sqrt{\mtrx{KPCP^TK^T}} = \sqrt{\mtrx{KgK^T}} \equiv \sqrt{\arbip{\mtrx{K}}{\mtrx{K}}} \equiv ||\mtrx{K}||.
\end{equation*}
If we let $\mtrx{S^T} = \mtrx{g}^{-1}\mtrx{PQ}$ then we find that the numerator
of the SNR is the inner product of $\mtrx{K}$ and $\mtrx{S}$:
\begin{equation*}
\mtrx{KPQ} = \mtrx{KgS^T} = \arbip{\mtrx{S}}{\mtrx{K}}.
\end{equation*}
The SNR is therefore at a maximum when $\mtrx{K}$ and $\mtrx{S}$ are aligned:
\begin{equation*}
\rho = \max_{\theta,\phi,\kappa} \frac{\arbip{\mtrx{S}}{\mtrx{K}}}{||\mtrx{K}||} = \max_{\theta,\phi} ||\mtrx{S}||.
\end{equation*}
The magnitude of $\mtrx{S}$ is:
\begin{equation*}
||\mtrx{S}|| = \sqrt{\mtrx{SgS^T}} = \sqrt{\mtrx{Q^T P^T}(\mtrx{g}^{-1})^\mtrx{T}\mtrx{g}\mtrx{g}^{-1}\mtrx{PQ}}.
\end{equation*}
Since $\mtrx{g}$ is symmetric its inverse is also symmetric. The SNR is thus:
\begin{equation}
\label{eqn:SNRmaxK}
\rho = 
    \begin{cases}
    \max_{\phi} P_{1i} Q_i \left(P_{1j} C_{jk} P_{k1}\right)^{-1/2}              & \text{if $\theta = \pi/2$}, \\
    \max_{\theta,\phi} \sqrt{\mtrx{Q^T P^T}(\mtrx{PCP^T})^{-1}\mtrx{P Q}}   & \text{otherwise.}
    \end{cases}
\end{equation}
Note that in performing the $\kappa$ maximization we did not need to invoke the
non-spinning assumption. Thus the argument of Eq. \eqref{eqn:SNRmaxK} is valid
for all systems; if there is spin, then further maximizations would need to be
carried out over the spin components in addition to $\theta$ and $\phi$.

\section{Estimating false-alarm probability from a bank of templates}
\label{appdx:bankFAP}

The false-alarm probability of a template bank as a function of SNR depends on
the size of the parameter space being searched over. The larger the parameter
space, the greater the probability of getting a false alarm. Let us assume that
every template in the bank is independent of each other. As described in Sec.
\ref{sec:background}, in a search, intrinsic parameters are maximized over by
selecting the template that yields the largest SNR. Due to this maximization,
only one template can produce a trigger at any given time, making the
probability of getting a trigger with some SNR from each template mutually
exclusive.  The false-alarm probability of the bank is thus:
\begin{equation*}
\mathcal{F}(\rho) = \sum_{k=1}^{N} \int_{\rho}^{\infty}\pdf{B}(\rho'| k)\d \rho',
\end{equation*}
where $N$ is the number of templates in the bank and $\pdf{B}(\rho' | k)$ is
the probability density function of the SNR of each template after maximization. Due to
the bank maximization $\pdf{B}(\rho | k)$ is given by the probability that the
template produces an SNR equal to $\rho$ times the probability that every other
template produces an SNR less than $\rho$; thus:
\begin{align}
\label{eqn:bankFAP_general}
\mathcal{F}(\rho) &= \sum_{k=1}^N \int_{\rho}^{\infty} \pdf{P}(\rho'|k) \prod_{l \neq k}^{N} \left[\int_{0}^{\rho'} \pdf{P}(\rho''| l)\d \rho'' \right] \d \rho' \nonumber \\
    &= \sum_{i=1}^N \int_{\rho}^{\infty} \pdf{P}(\rho'|k) \prod_{l \neq k}^N \cdf{P}(\rho'|l) \d \rho'.
\end{align}
Here $\pdf{P}(\rho|k)$ is the SNR distribution in noise of the $k^\th$ template
and $\cdf{P}(\rho | l)$ is the cumulative distribution function of the $l^\th$
template. 

In Gaussian noise, every dominant-mode template is $\chi$ distributed with two
degrees of freedom. For a dominant-mode bank Eq. \eqref{eqn:bankFAP_general}
simplifies to:
\begin{equation}
\label{eqn:bankFAP_chi2}
\mathcal{F}_\dm(\rho) = 1 - \left(1 - e^{-\frac{\rho^2}{2}}\right)^N.
\end{equation}
In our case, $N = 19\,800$ and $\rho = 8$; the probability that the
dominant-mode bank produces a false alarm at SNR 8 is therefore $\approx 2.5
\times 10^{-10}$.

Note that Eq. \eqref{eqn:bankFAP_chi2} approaches $1$ as $N \rightarrow \infty$
for all $\rho > 0$. This is due to our assumption that the templates are
independent of each other.  We expect that as we increase the minimal match of
the bank to $100\%$ (which would require an infinite number of templates) the
false-alarm probability would instead approach some limiting value for a given
$\rho$. The minimal match of our template bank is $97\%$; assuming that
templates are independent might therefore seem exceedingly na\"ive. To test
this assumption we filtered the dominant mode template bank in $10\,000$
realizations of noise. In each realization we selected the template with the
best SNR to provide a measure of false-alarm probability as a function of SNR.
We found good agreement between these results and Eq.
\eqref{eqn:bankFAP_chi2}. Thus, despite a seemingly large overlap between
neighboring templates, our assumption of independence appears to be
approximately valid for the minimal match of our bank.

To model a search with sub-dominant modes we assume a bank with the same number
of templates and with the same $(M, q)$ coordinates as the dominant-mode bank.
In such a search we would maximize $\rho$ over $\theta, ~\phi$ and $\kappa$ for
each template, then maximize over the entire bank. To find $\mathcal{F}(\rho)$
of this bank we need the probability density function of the SNR for each
template in stationary Gaussian noise, $\pdf{P}(\rho|k)$. To find that we need
an expression for the maximized SNR when sub-dominant modes are included. In
Appendix \ref{appdx:maxHMsnr-analytic} we analytically maximize $\rho$ over
$\kappa$ to get Eq. \eqref{eqn:SNRmaxK}. This equation depends on the matrix
$\mtrx{Q}$, the elements of which are the overlap between the template and the
detector data [see Eq. \eqref{eqn:QmatrixDef}]. In stationary Gaussian noise the
$Q_i$ are Gaussian random variables with variance $\sigma_i^2 = \sum_j C_{ij}$,
where $C_{ij}$ are the elements of the covariance matrix defined in Eq.
\eqref{eqn:covar_matrix}. To find $\pdf{P}(\rho|k)$ from this multivariate
Gaussian distribution we need to maximize over $\theta$ and $\phi$. This
maximization is not trivial, however, and so we must find the SNR distribution
numerically.

One way to find $\pdf{P}(\rho|k)$ is to generate many realizations of noise,
filter it to get $\mtrx{Q}$, then perform the maximization in Eq.
\eqref{eqn:SNRmaxK} for each template. However, we expect the probability of
getting $\rho \approx 8$ to be extremely small: the probability of getting
$8\pm 0.1$ from a $\chi$ distribution with 10 degrees of freedom (the upper
limit of what we expect from a template with 5 modes) is order $10^{-9}$.
Getting an accurate measure of $\pdf{P}(\rho|k)$ around SNR $8$ is thus
computationally intractable using this ``brute force" method. Instead, we
follow a procedure similar to that used in Ref. \cite{pan:2004} to find the
probability density function.

To simulate a particular realization of $\mtrx{Q}$ we do not need to do any
matched filtering; instead we draw a set $\mtrx{Z}$ of pseudo-random values
from a Gaussian distribution with zero mean and unit variance. $\mtrx{Q}$ is
then\footnote{Since $\mtrx{C}$ is positive-definite $\sqrt{\mtrx{C}}$ is real;
we find it from the eigendecomposition of $\mtrx{C}$. Specifically,
$\sqrt{\mtrx{C}} = \mtrx{T}\sqrt{\mtrx{\Lambda}}\mtrx{T}^{-1}$ where $\mtrx{T}$
is the matrix of the eigenvectors of $\mtrx{C}$ and $\mtrx{\Lambda}$ is the
diagonal matrix formed from the eigenvalues.}:
\begin{equation*}
Q_i = \sqrt{C_{ij}} Z_j.
\end{equation*}
We think of $\mtrx{Q}$ as being a vector in a $\nu$ dimensional space
$\mathcal{S}$, where $\nu$ is equal to twice the number of modes; the relative
size of each dimension is determined by the covariance matrix.

\begin{figure}
\includegraphics[width=\columnwidth]{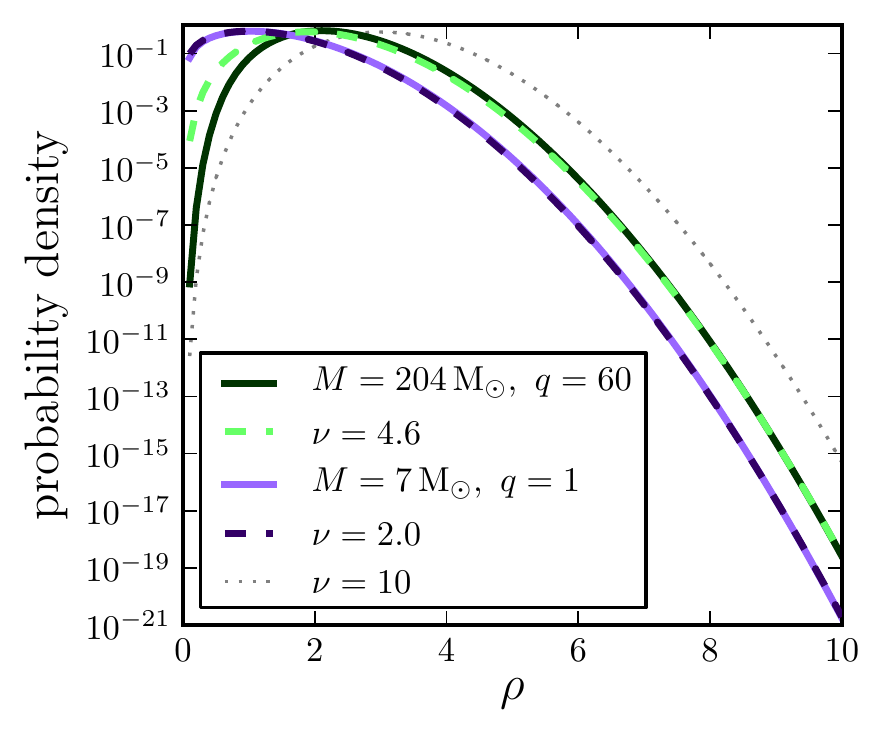} \\
\includegraphics[width=\columnwidth]{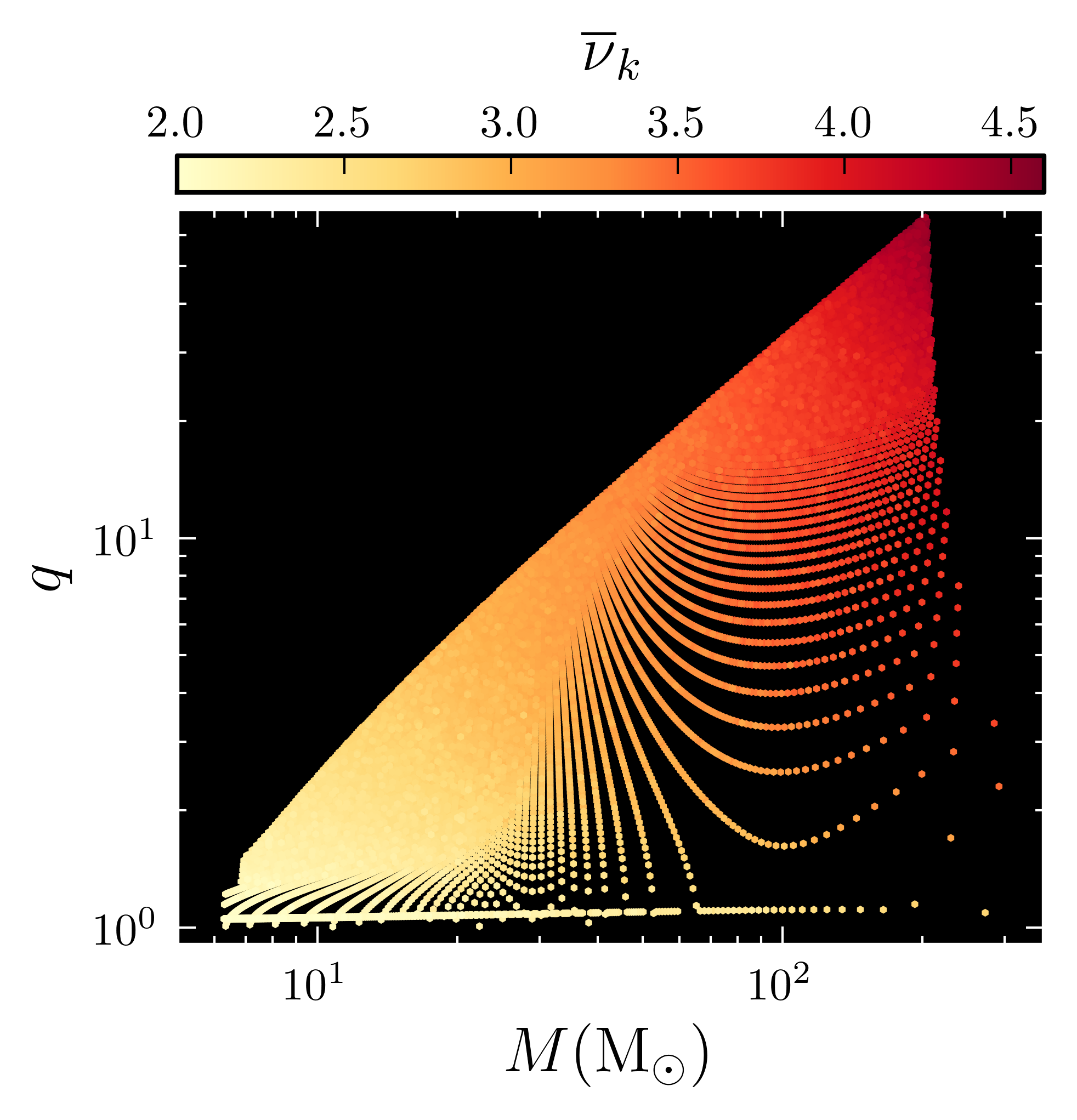}

\caption{SNR distribution in noise of two templates with sub-dominant modes
($\pdf{P}(\rho|k)$) versus SNR (top) and fitted number of degrees of freedom
$\overline{\nu}_k$ for each template in total mass and mass-ratio $q$ (bottom).
The dashed lines in the top plot show a $\chi$ distribution with the fitted
$\overline{\nu}_k$ to $\pdf{P}(\rho|k)$ of the templates shown. Fits were done
between $\rho = 7$ and 9.}

\label{fig:HMbankNdofs}
\end{figure}

Now let $r = ||\mtrx{Z}||$ and $||\mtrx{\hat{Z}}|| = 1$ such that $\mtrx{Z} = r
\mtrx{\hat{Z}}$. Define $\hat{\rho}$ to be the maximized SNR we obtain from a
realization of $\mtrx{\hat{Q}} = \sqrt{\mtrx{C}}\mtrx{\hat{Z}}$. From From Eq.
\eqref{eqn:SNRmaxK} we see that:
\begin{equation}
\label{eqn:rho_r_dep}
\rho(\mtrx{Q}, k) = r\hat{\rho}(\mtrx{\hat{Q}}, k) \equiv r\hat{\rho}(\Omega, k).
\end{equation}
Here, $\Omega$ is the solid angle describing the direction of $\mtrx{Q}$ in
$\mathcal{S}$, and we have made the dependence of the SNR on the intrinsic
parameters of the $k^\th$ template explicit. Since each element in $\mtrx{Z}$
is a Gaussian random variable with zero mean and unit variance, $r$ --- which
is the quadrature sum of these variables --- is $\chi$ distributed with $\nu$
degrees of freedom:
\begin{equation}
\pdf{R}(r| \nu) = \frac{r^{\nu-1}e^{-r^2/2}}{2^{\nu/2 -1}\Gamma(\nu/2)}.
\end{equation}
The number of degrees of freedom is equal to the dimension of $\mtrx{Q}$, which
is twice the number of modes. In our case, $\nu = 10$.

Using the coordinate transformation given by Eq. \eqref{eqn:rho_r_dep} we find:
\begin{equation}
\pdf{P}(\rho| k, \Omega, \nu) = \frac{1}{\hat{\rho}(\Omega, k)} \pdf{R}\left(\left.\frac{\rho}{\hat{\rho}(\Omega, k)}\right| \nu \right).
\end{equation}
Marginalizing out $\Omega$ yields $\pdf{P}(\rho| k, \nu)$:
\begin{equation}
\label{eqn:intHMtemplatePDF}
\pdf{P}(\rho|k, \nu) = \left. \int \frac{1}{\hat{\rho}(\Omega, k)} \pdf{R}\left(\left.\frac{\rho}{\hat{\rho}(\Omega, k)}\right| \nu\right) \d\Omega \right/ \int \d \Omega.
\end{equation}
We solve this via Monte Carlo integration. For each point in the Monte Carlo we
generate a normalized random vector $\mtrx{\hat{Z}}$. We use Eq.
\eqref{eqn:SNRmaxK} to find $\hat{\rho}(\Omega)$ for this realization, with
$\mtrx{\hat{Q}} = \sqrt{\mtrx{C}}\mtrx{\hat{Z}}$. We then find $\pdf{P}(\rho|
k, \nu)$ with $\nu = 10$ for several different values of $\rho$, terminating
the Monte Carlo when the error on $\pdf{P}(\rho|k)$ at $\rho = 9$ is less than
$20\%$. 

The top plot in Fig. \ref{fig:HMbankNdofs} shows $\pdf{P}(\rho|k)$ for two
different templates. We find that for $\rho \gtrapprox 7$, $\pdf{P}(\rho|k)$
approaches the $\chi$ distribution with non-integer number of degrees of
freedom $\pdf{X}(\rho| \nu_k)$ for all templates. We therefore fit
$\pdf{X}(\rho| \nu_k)$ to $\pdf{P}(\rho|k)$ between $\rho = 7$ and $9$ by
maximizing over $\nu_k$. The bottom plot in Figure \ref{fig:HMbankNdofs} shows
the best-fit $\nu_k$ ($\overline{\nu}_k$) for each template in the bank. If all
of the modes were independent of each other $\overline{\nu}_k$ would be equal
to $10$ for all $k$. Instead we find that the largest $\overline{\nu}_k \approx
4.6$, which occurs at the largest mass-ratio and total mass part of the bank.
As the templates approach the equal mass line the number of degrees of freedom
approaches two. This is expected: as $q \rightarrow 1$, the sub-dominant modes
become small relative to the dominant mode, and $\pdf{P}(\rho|k)$ reduces to a
$\chi$ distribution with 2 degrees of freedom.

To solve Eq. \eqref{eqn:bankFAP_general} we substitute $\pdf{X}(\rho|
\overline{\nu}_{k})$ and $\cdf{X}(\rho|\overline{\nu}_k)$ for $\pdf{P}(\rho|k)$
and $\cdf{P}(\rho|k)$, then numerically integrate for several different values
of $\rho$. Inverting yields $\rho(\mathcal{F})$, which we solve for
$\mathcal{F} = \mathcal{F}_\dm(\rho = 8)$.

\bibliography{references}

\end{document}